\documentclass[prd,showpacs,twocolumn,superscriptaddress,amssymb,amsmath,nofootinbib]{revtex4}
\usepackage{graphicx}
\usepackage{subfigure}

\newcommand\etal{{\it et al. }}

\newcommand\wtilde{\widetilde}
\newcommand\dd{\partial}
\newcommand\dr{\mathrm{d}}
\newcommand\ra{\rightarrow}
\newcommand\bb[1] {   \mbox{\boldmath{$#1$}}  }
\newcommand\del{\bb{\nabla}}

\newcommand\zobs{z_{\text{obs}}}
\newcommand\zfields{z_{\text{fields}}}
\newcommand\zform{z_{\text{form}}}
\newcommand\deltac{\delta_{\text{crit,i}}}
\newcommand\hmpc{\,h^{-1}\,\text{Mpc}}
\newcommand\hovermpc{\,h\,\text{Mpc}^{-1}}
\newcommand\hmsun{\,h^{-1}\,\text{M}_{\odot}}
\newcommand\lsim{\mathrel{\rlap{\lower4pt\hbox{\hskip1pt$\sim$}}
        \raise1pt\hbox{$<$}}}
\newcommand\gsim{\mathrel{\rlap{\lower4pt\hbox{\hskip1pt$\sim$}}
        \raise1pt\hbox{$>$}}}

\begin{document}
\title{Scale-dependent halo bias from scale-dependent growth}
\date{\today}

\author{Kyle Parfrey}
\email{kyle@astro.columbia.edu}
\affiliation{Department of Astronomy, Columbia University, New York, New York 10027}
\affiliation{Institute for Strings, Cosmology, and Astroparticle Physics (ISCAP),
Columbia University, New York, New York 10027}
\author{Lam Hui}
\email{lhui@astro.columbia.edu}
\affiliation{Institute for Strings, Cosmology, and Astroparticle Physics (ISCAP),
Columbia University, New York, New York 10027}
\affiliation{Department of Physics, Columbia University, New York, New York 10027}
\author{Ravi K. Sheth}
\email{shethrk@physics.upenn.edu}
\affiliation{
Center for Particle Cosmology, University of Pennsylvania,
209 South 33rd Street, Philadelphia, PA 12104
}

\begin{abstract}
We derive a general expression for the large-scale halo bias, in
theories with a scale-dependent linear growth, using the excursion
set formalism. Such theories include modified gravity models, and
models in which the dark energy clustering is non-negligible.
A scale dependence is imprinted in both the formation and evolved biases
by the scale-dependent growth. Mergers are accounted for in our derivation, which thus
extends earlier work which focused on passive evolution.
There is a simple analytic form for the bias for those theories in which the nonlinear collapse of perturbations is 
approximately the same as in general relativity.
As an illustration, 
we apply our results to a simple Yukawa modification of gravity, 
and use SDSS measurements of the clustering of luminous red galaxies 
to constrain the theory's parameters.

\end{abstract}

\maketitle

\section{\label{sec:intro} Introduction}

The cosmological constant ($\Lambda$) + cold dark matter (CDM)
+ general relativity (GR) model
has been very successful in accounting
for current cosmological data.
It is incumbent upon us to put this standard model to further tests.
One of its characteristic predictions is a scale-independent sub-Hubble
linear growth, which can be violated if dark energy clusters\footnote{Dark energy other than the cosmological constant generally clusters,
though the degree of sub-Hubble clustering is often negligible. Exceptions
include models where the dark energy sound speed is substantially less
than unity, e.g.~\cite{Creminelli2009}.
}, or if gravity is modified
\cite{Dvali:2000hr,Nicolis2009,deRham2010,Fierz1956,Jordan1959,Brans1961,Buchdahl:1983zz,Carroll2004}.
Both types of models introduce new scales into the growth of structure: the Jeans
scale in the case of clustered dark energy, and 
the GR-to-non-GR transition scale in the case of modified gravity.
The most direct way to test for this effect is to measure
the matter power spectrum at different redshifts, and 
reconstruct the growth factor as a function of scale. 
Here, we focus on a corollary of a scale-dependent growth, a scale-dependent
halo bias, which in principle allows us to discern scale dependence even
with measurements of the large-scale structure at a single redshift.

The fact that a scale dependence in growth implies scale dependence
in halo bias (on linear scales) was pointed out by \cite{Hui:2007zh}
(henceforth HP), generalizing earlier work by
\cite{Fry:1996fg,Tegmark1998}:
\begin{eqnarray}
\label{HPresult}
b_1 (k_0, z_{\rm obs}; z_{\rm form}) 
= 1 + [b_1 (z_{\rm form}) - 1] {D (k_0, z_{\rm form}) \over
D (k_0, z_{\rm obs})}
\end{eqnarray}
where $b_1(k_0, z_{\rm obs}; z_{\rm form})$ signifies the linear bias
on scale $k_0$ observed at redshift $z_{\rm obs}$, for
haloes that form at redshift $z_{\rm form}$.
The symbol $D$ denotes the linear growth factor at
the relevant scale and redshift. 
This expression, which assumes passive evolution, i.e.
halo number conservation after formation, 
tells us that the observed bias would inherit a scale
dependence from the growth, even if the formation bias is scale-independent.
In this paper, we wish to relax these two assumptions:
scale-independent formation bias, and no mergers.

The paper is organized as follows.
The extended Press-Schechter, or excursion set, formalism 
\cite{Press:1973iz, Bardeen:1986, Bond:1990iw, Lacey:1993, Sheth:1998ew} 
is described and generalized to allow for
a scale-dependent growth in Sec.~\ref{sec:exset}. This is used to
compute the halo mass function and the halo bias. 
In Sec.~\ref{sec:illustration}, we describe
an illustrative example, a modified gravity model of the Yukawa type,
and present a calculation of the linear growth factor and the excursion
barrier (collapse threshold).
We present the halo bias for this example, and
compare it with observations.
We conclude in Sec.~\ref{sec:conclusions}. 
In this paper, for the purpose of illustration, we have chosen to focus
on one particular model of modified gravity. 
Our results in \S \ref{sec:exset} for the scale-dependent halo bias
are, on the other hand, fairly general. A recipe for using
these results in more general settings is summarized in 
Sec. \ref{sec:conclusions}. Readers who are interested primarily
in applications, and not on the derivation, can skip directly to
Sec. \ref{sec:conclusions}.
In Appendix~\ref{sec:extrap} we discuss some details concerning the use of the 
excursion set method for theories with a scale-dependent growth factor.
In Appendix \ref{sec:scalartensor} we give the derivation of the Yukawa model 
from a scalar-tensor theory, and
we describe our spherical collapse model
in Appendix~\ref{sec:SC}.

Before we proceed, let us briefly discuss the connection with
some of the literature on the subject.
The halo mass function for a Yukawa theory like the one we study
was computed by \cite{Martino:2009}. Our paper follows their formalism,
and it is in a sense a straightforward extension to compute the conditional
mass function and halo bias.
Halo bias in $f(R)$ gravity and the DGP models has been measured from
numerical simulations in
\cite{Schmidt:2009,Khoury2009,Schmidt2009dgp,Chan2009,Schmidt:2010} .
They find fair agreement between simulations and the bias derived from
a modified Sheth and Tormen \cite{Sheth:1999mn} mass function 
whose parameters reflect the altered spherical collapse.

There is a large literature on testing GR using the growth of 
large-scale structure, e.g. \cite{Lue2003}.
Most studies allow for a scale-dependent growth factor, but
ignore its effect on the galaxy bias. Our expression for the bias
should be useful for incorporating the latter effect into such studies.

There is also a substantial literature on a large-scale
scale-dependent bias from primordial non-Gaussianity,
e.g. \cite{Dalal2008}.
This other source of scale dependence should be distinguishable
from that from growth; we will discuss this in 
Sec. \ref{sec:conclusions}.

\section{\label{sec:exset}Excursion set theory of haloes with a scale-dependent growth factor}

We describe in detail here the excursion set formalism. 
Much of the discussion replicates the standard
treatment, but with special care taken to allow for a scale-dependent growth factor.

\subsection{\label{sec:randomwalk} Random walks}

\begin{figure}
\includegraphics[width=8.6cm]{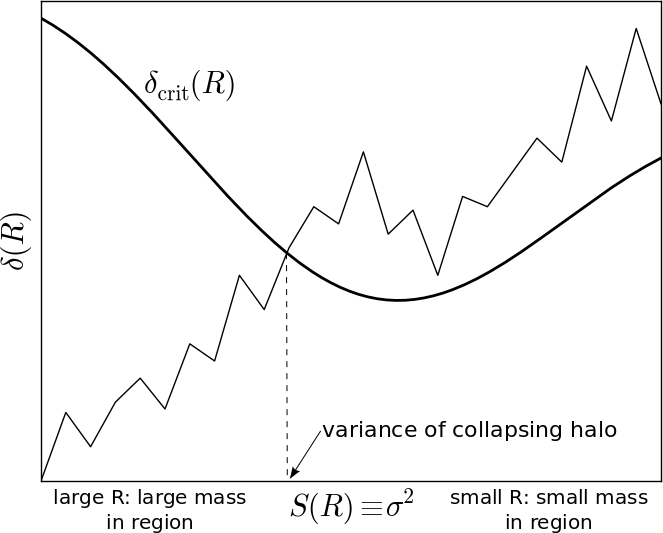}
\caption{\label{fig:randomwalk} 
Random walks: notation for the 
excursion set model. A point is contained in a
halo corresponding to the first upcrossing of the barrier, here $\delta_{\rm crit}$, 
by the random walk.}
\end{figure}

The premise of the excursion set theory is that a halo will form from a region
of a certain size when the overdensity of matter $\delta$, smoothed over that
region, is greater than a critical value $\delta_{\rm crit}$ 
\cite{Press:1973iz}. 
In FIG.~\ref{fig:randomwalk}, the mass overdensity, at
a point, smoothed on a comoving 
scale $R$, is plotted against the variance $S$ on that scale around
that point, where
\begin{equation}
S(R) = \int_0^{\infty} \frac{\mathrm{d}^3k}{(2\pi)^3} P(k) |\wtilde{W}(k R)|^2\,,
\label{eq:S}
\end{equation}
$P(k)$ is the matter power spectrum, and $\wtilde{W}(kR)$ the Fourier-space window function.
As the smoothing scale is decreased, the smoothed density field undergoes
a random walk. 
The halo mass $M$ is given by
$M = (4\pi/3)\bar{\rho}(z=0) R^3$, where
$\bar\rho$ is the mean matter density today. Hence, 
$M$, $S$ and $R$ can all be thought of as equivalent
variables.

Following common practice, we smooth the density field with a 
real-space top-hat filter
(even though, strictly speaking, a true random walk requires
a top-hat filter in Fourier space \cite{Bond:1990iw}).
In an important sense all 
of our analysis is `native' to real space; for all quantities not explicitly
marked as being in Fourier space by a $\wtilde{\;\,}$, we define the 
wavenumber
associated with
a scale $R$ as simply $k \equiv 1/R$. 
The Fourier transform of a real-space top hat is
\[
\wtilde{W}(kR) = \frac{3}{x^3} \left\{ \sin(x) - x\cos(x)  \right\} \,,\; x\equiv k R\,.
\]

Following \cite{Martino:2009}, this random walk exercise
is carried out at some initial redshift $z_i$, 
chosen to be early enough that 
the linear growth is scale-independent
(i.e. neither dark energy nor modified gravity is important
by this redshift\footnote{
In the standard model of GR + $\Lambda$ or (nearly) homogeneous dark energy, 
the linear growth is scale-independent, the redshift at which
the density field is defined and the random walk is carried out is 
immaterial. This is discussed in Appendix \ref{sec:extrap}.
}).
Throughout this paper we set $z_i = 100$, although its
precise value is not important as long as it is sufficiently early.
We add the subscript $i$ to 
the critical threshold for halo formation
$\delta_{\rm crit, i}$ to remind ourselves that it is the comparison
between the smoothed $\delta_m$ at $z_i$ against this threshold that
defines haloes. The value for $\delta_{\rm crit, i}$
in general depends on both the halo mass of interest (i.e. the
barrier is not necessarily flat, unlike in the standard model), 
and the actual physical redshift of halo formation $z_{\rm form}$.

\subsection{\label{sec:massfn} Unconditional first-crossing distribution \& the mass function}

The central quantity in the excursion set theory is the first-crossing distribution, $f$.
Define $f(S_1,\delta_1|0,0)\,\mathrm{d}S_1$ as the probability that a random walk starting
at the origin will first cross the barrier, of height $\delta_1 \equiv \deltac$, 
at $S = S_1 \pm \mathrm{d}S_1/2$. This is equivalent to saying that 
$f(S_1,\delta_1|0,0)|\mathrm{d}S_1/\mathrm{d}M_1|\,\mathrm{d}M_1$ is the probability 
that 
our point (around which this smoothing or random walk exercise
is performed) belongs to a halo of mass $M_1 \pm \mathrm{d}M_1/2$. We can think of 
$(S_1,\delta_1)$ as the target point we are aiming for on our random walk.
 
To find $f$, it is sometimes convenient to
consider first the cumulative crossing distribution, $F$, which is the 
probability that the walk first crosses the barrier at $S<S_1$:
\[
F(S_1,\delta_1|0,0) = \int_0^{S_1} f(S,\delta|0,0)\,\dr S\,,
\]
where $\delta$ is the critical threshold $\deltac$,
generally a function of $S$; in the case where it is $S$-independent,
i.e. flat barrier, $F$ (and therefore $f=\partial F/\partial S_1$)
can be found using a clever trick due to Chandrasekhar 
\cite{Chandrasekhar:1943ws}, 
or by modelling the random walk as a diffusive process 
(see \cite{Bond:1990iw,Lacey:1993}). 
For a more general barrier, an alternative method for finding $f$ will
be discussed in \S \ref{sec:calcf}.
We assume that the probability distribution 
of the density field as a function of the smoothing scale 
$R$ is Gaussian:
\[
P_G(\delta, S(R)) d\delta = \frac{1}{\sqrt{2\pi S(R)}}\exp{[-\delta^2/2 S(R)]}\,\dr \delta\,.
\]

Using the above, we see that the
mass function of haloes (the physical number density of 
haloes as a function of mass, smoothed over the whole universe) which are
\emph{identified} at $\zform$ and \emph{observed} at $\zobs$ is given by
\begin{equation}
n(M_1)\dr M_1 = \frac{\bar{\rho}(\zobs)}{M_1} f(S_1,\delta_1|0,0)\left|\frac{\dr S_1}{\dr M_1}\right|\dr M_1\, ,
\label{eq:massfn}
\end{equation}
where $\bar\rho(\zobs)$ is the mean matter density at $\zobs$.
The mass function has an implicit dependence on $\zform$ through the
dependence of the threshold/barrier $\delta_1$ on the formation redshift.

\subsection{\label{sec:bias} Conditional first-crossing distribution \& halo bias}
\subsubsection{Lagrangian space}
We can now move the starting point of our random walk away from the origin, to some other
point on the $S$--$\delta$ plane. Call this point $(S_0,\delta_0)$. This is equivalent to
calculating all our quantities within a region of the size on which the mass variance is
$S_0$, and which has a mass overdensity $\delta_0$. The differential 
probability that a random walk starting from this point will first 
cross the barrier $\delta_1$ is $f(S_1,\delta_1|S_0,\delta_0)$. In the flat-barrier case
this only depends on $S_1 - S_0$ and $\delta_1 - \delta_0$.

Using this conditional first-crossing distribution we can construct an expression for the
overdensity of haloes within a region of
$(S_0, \delta_0)$ \cite{Mo:1995cs,Sheth:1998xe}. 
Just as we derived the unconditional mass function
above, we can derive a conditional mass function. We can see from our definitions that the 
mean number $\mathcal{N}$ of $M_1$ haloes in a region of total mass $M_0$ is  
\begin{equation}
\mathcal{N}(M_1,\delta_1|M_0,\delta_0) = \frac{M_0}{M_1} f(S_1,\delta_1|S_0,\delta_0) \left|\frac{\dr S_1}{\dr M_1}\right|\,,
\label{eq:N}
\end{equation}
and so their number density is $\mathcal{N}/V_0$ if the region has volume $V_0(M_0)$. 

We define the overdensity of haloes `1' in a region `0' as the fractional overabundance
of these haloes in this region compared to their number density smoothed over the
whole universe, giving
\begin{equation}
\delta_h^L(1|0) = \frac{\mathcal{N}(M_1,\delta_1|M_0,\delta_0)}{n(M_1,\delta_1)V_0} - 1\,.
\label{eq:lbias}
\end{equation}
This is the halo overdensity in Lagrangian co-ordinates, in which 
$M_0 \approx \bar{\rho}(z_i)V_0$, defining $V_0$.
Indeed, all quantities in this Lagrangian space discussion,
including $\delta_0, V_0, \delta_1$ and so on, are defined at 
the initial redshift $z_i$.

\subsubsection{Eulerian space}
In Eulerian co-ordinates we need to replace $V_0$ with the evolved Eulerian volume. In the
above, the Lagrangian space quantities $V_0$ and $\delta_0$ are calculated at 
$z_i$. We now have three redshifts: the initial 
redshift $z_i$ (which we always set to 100), the identification redshift 
of the haloes $\zform$ (in a sense, their formation redshift), and the redshift at
 which the haloes are observed $\zobs$.

Label as $\delta$ the physical matter overdensity of the region of interest, 
at the observation redshift $\zobs$. It will be related to the Lagrangian space $\delta_0$ by
\begin{equation}
\delta = \frac{D(S_0,\zobs)}{D(S_0,z_i)}\,\delta_0\,,
\label{eq:delta}
\end{equation}
because $\delta_0$ is defined at $z_i$. 
The use of the linear growth factor $D$ is justified for
a small $S_0$, or correspondingly a large region.
By the time of observation, the volume of the
observed patch will have contracted from its Lagrangian-space value, because conservation
of mass requires that the volume decreases as $\delta$ increases. Label as $V$ the volume
of our region at $z_i$, taking into account the contraction it will experience by the
time it reaches $\zobs$. Since $\delta(z_i) \approx 0$, we find
\[
M_0 \approx \bar{\rho}(z_i) V_0 \equiv \bar{\rho}(z_i)(1+\delta) V
\quad \longrightarrow \quad V = \frac{V_0}{1+\delta}\,.
\]

Substituting $V_0 \ra V$ into Eq.~(\ref{eq:lbias}), and using Eqs.~(\ref{eq:massfn}) and
(\ref{eq:N}), gives the Eulerian halo overdensity $\delta_h^E$:
\begin{equation}
\delta_h^E = (1 + \delta)\frac{f(S_1,\delta_1|S_0,\delta_0(\delta))}{f(S_1,\delta_1|0,0)} - 1\,.
\label{eq:delta_h}
\end{equation}

It is clear that $\delta_h$ (dropping the `E' from now on) will in general not be the same
as the matter overdensity in the same region, $\delta$. A positive $\delta$ will usually help
haloes reach the collapse threshold and increase their overdensity, with the opposite 
effect for negative $\delta$. There are also halo exclusion effects; for example, the 
halo density will be suppressed unless their mass is much smaller 
than the mass available in the region. From the random walk perspective, 
the `time' available for 
barrier crossing is $S_1 - S_0$, in which time the walk must 
travel a `distance' $\delta_1 - \delta_0$. You can suppress halo formation either by
reducing the time available ($M_{\text{halo}} \equiv M_1$ 
approaching $M_0$) or by increasing the required
distance (increasing the height of the barrier).  

Since we know $\delta_h$ is a function of $\delta$, we can expand it
as a Taylor series \cite{Fry1992},
\begin{eqnarray}
\label{FG}
\delta_h = \sum_{k=0}^{\infty} \frac{1}{k!} b_k \delta^k\,.
\end{eqnarray}
The $b_k$ co-efficients are the bias parameters. The first, $b_0$, 
vanishes in the limit of $S_0 \ll S_1$. Here, we are primarily
concerned with $b_1$, the linear bias, 
in the $\delta \ra 0$ limit (because we are 
interested in large regions, on which scales the mass overdensity is 
expected to be small). The linear bias in the limit $\delta \ra 0$ is just
 $\dd \delta_h/\dd \delta$ evaluated at $\delta = 0$, 
and so, using $\dd \delta_0/\dd \delta = D(S_0,z_i)/D(S_0,\zobs)$, we find
\begin{align}
b_1(\delta \ra 0) &= \frac{1}{f(S_1,\delta_1|0,0)} \Biggl\{ f(S_1,\delta_1|S_0,0) \nonumber\\*
 &\quad + \frac{D(S_0,z_i)}{D(S_0,\zobs)}\left. \frac{\dd f(S_1,\delta_1|S_0,\delta_0)}{\dd \delta_0}\right |_{\delta_0 = 0}  \Biggr\}\,.
\label{eq:b}
\end{align}

Recall that $S_0$, $R_0$, and $k_0$ are all equivalent variables; $S_0$ and $R_0$ are related 
by Eq.~(\ref{eq:S}), and $k_0 \equiv 1/R_0$.
The $D(S_0,z_i)/D(S_0,\zobs)$ factor in this equation will make 
the large-scale bias scale-dependent in many modified gravity theories, or theories in
which dark energy clusters in a non-negligible manner.

\subsection{\label{sec:flatbarrier} Flat barrier approximation}

As we will see in \S \ref{sec:illustration}, a flat barrier may be a good 
approximation for some theories. 
This is not surprising even for non-standard theories: haloes of a sufficiently
small mass cross the barrier in a regime where essentially the standard
story applies, i.e. the crossing occurs on small scales where
gravity is Newtonian and dark energy typically is quite smooth.
A useful analytic form for the halo bias can be written down in this case.
The conditional first crossing distribution is \cite{Bond:1990iw}
\[
f(S_1,\delta_1|S_0,\delta_0) = \frac{1}{\sqrt{2\pi}}\frac{\delta_1-\delta_0}{(S_1-S_0)^{3/2}}
\exp{\left[-\frac{(\delta_1-\delta_0)^2}{2(S_1-S_0)}\right]}\,,
\]
and so, using Eq.~(\ref{eq:b}), we find the Eulerian-space bias to be
\begin{align}
b_1 &= \left(\frac{S_1}{S_1-S_0}\right)^{3/2}\left\{1 + \frac{D(S_0,z_i)}
{D
(S_0,\zobs)}\left(\frac{\delta_1}{S_1-S_0} - \frac{1}{\delta_1}\right)\right\}\nonumber\\*
&\quad \times\exp{\left[-\frac{\delta_1^2
S_0}{2 S_1(S_1-S_0)} \right]}\,.
\label{eq:bsimplefull}
\end{align}

We are interested in scales on which
$S_0 \ll S_1$, i.e. the halo mass of interest
is much smaller than the mass encompassed in
the region over which we are calculating the
clustering. Therefore,
\begin{equation}
b_1(S_0, z_{\rm obs}; z_{\rm form}) = 1 + \frac{D(S_0,z_i)}{D(S_0,\zobs)}\left(
\frac{ {\delta_1}^2}{S_1} - 1\right)
\frac{1}{\delta_1}  \, .
\label{eq:bsimple}
\end{equation}
The bias $b_1$ depends on $z_{\rm form}$
because $\delta_1$ is the initial overdensity threshold for collapse
at $z_{\rm form}$. 

As a check, it can be seen this reduces to the
Mo \& White \cite{Mo:1995cs} result in the standard model, by recognizing that
both $\delta_1$ and $S_1$ in our description are defined
at the initial redshift $z_i$ --- one can use part of
the factor $D(S_0, z_i)/D(S_0, z_{\rm obs}) = 
[D(z_i)/D(z_{\rm form})]
[D(z_{\rm form})/D(z_{\rm obs})]$ to rescale ${\delta_1}^2/S_1$
and $\delta_1$ down to $z_{\rm form}$ (recalling that
$D$ is scale-independent in the standard model), and obtain
$b_1 = 1 + [D(z_{\rm form})/D(z_{\rm obs})](\nu^2 - 1)/\delta_{c}$, where 
$\delta_{c}$ is the linearly extrapolated overdensity of collapse
at $z_{\rm form}$\footnote{
\label{1686}
For instance, 
$\delta_{c} \sim 1.686$ for a universe with 
a critical matter density and no cosmological constant
or dark energy.
More generally, for a flat universe with matter
and $\Lambda$, $\delta_c \sim 1.686
\times (1 + 0.123 {\,\rm log}_{10} \Omega_f)$,
with $\Omega_f = 
\Omega_m (1 + z_{\rm form})^3 / 
[\Omega_m (1 + z_{\rm form})^3 + \Omega_\Lambda]$
\cite{Kitayama1996}.
},
and
$\nu \equiv \delta_{c}
/\sqrt{S_{1, \rm form}}$ with $S_{1, \rm form}$ being 
the the variance at $z_{\rm form}$ for our halo mass of interest $M_1$.
The bias in its most familiar form obtains
when one sets $z_{\rm form} = z_{\rm obs}$.

In non-standard models, Eq. (\ref{eq:bsimple}) implies 
the linear bias $b_1$ is scale ($S_0$) dependent in general.
The scale dependence comes entirely
from the ratio of growth factor $D(S_0, z_i)/D(S_0, z_{\rm obs})$. 
The result resembles Eq. (\ref{HPresult}) \cite{Hui:2007zh}.
To facilitate comparison, it is helpful to rewrite 
Eq. (\ref{eq:bsimple}) as
\begin{equation}
b_1(S_0, z_{\rm obs}; z_{\rm form}) 
= 1 + (b_1(S_0, \zform) - 1)\frac{D(S_0,\zform)}{D(S_0,\zobs)},
\label{eq:oldbias}
\end{equation}
where $b_1(S_0, \zform)$ is the formation bias, given by
\begin{eqnarray}
\label{formationbias}
b_1(S_0, \zform) = 1 + 
\frac{D(S_0, z_i)}{D(S_0,\zform)}
\left ( \frac{{\delta_1}^2}{S_1} - 1\right)
\frac{1}{\delta_1} 
\end{eqnarray}
This is a trivial rewriting, but it makes clear that
the formation bias $b_1(S_0, \zform)$ is in general scale-dependent,
contrary to what was assumed in HP.
It also helps differentiate between two different effects:
one is the scale dependence at formation
(from $D(S_0, z_i)/D(S_0, \zform)$ in Eq. [\ref{formationbias}])
, and the other
is the scale dependence from passive evolution thereafter
(from $D(S_0, \zform)/D(S_0, \zobs)$ in Eq. [\ref{eq:oldbias}]).

Note that when $\zform \ne \zobs$, the haloes are identified
at $\zform > \zobs$, and maintain their identities until $\zobs$\footnote{These 
expressions work even for $\zform < \zobs$.
This is relevant if one identifies haloes at some
low redshift $\zform$, but is interested in the clustering of
the center of mass of their constituents at some earlier redshift $\zobs$.}.
Mergers in the form of accretion onto these haloes are allowed
between the two epochs, so long as the identification is unaltered,
i.e. the haloes, once identified, are conserved.
If this is not true --- if the haloes continue to merge with
each other all the way to $\zobs$ --- then one should use
$\zform = \zobs$. In this case, there is no
distinction between the bias observed at
$\zobs$ and the formation bias.

It is also instructive to further rewrite the formation bias
using the following relations:
\begin{subequations}
\begin{align}
\label{eq:deltazform}
\delta_1 = \delta_{1, {\rm form}}
\frac {D(S_1, z_i)}{D(S_1, \zform)} \, ,\\
S_1 = S_{1, {\rm form}} 
\frac {D^2(S_1, z_{i})}{D^2(S_1, \zform)} \, ,
\end{align}
\end{subequations}
where we have introduced $\delta_{1, {\rm form}}$
and $S_{1, {\rm form}}$ for the linear overdensity threshold
and the variance at $\zform$ (recall that our random
walks are performed at the initial redshift $z_i$,
hence $\delta_1$ and $S_1$ are defined then).
The formation bias can thus be written as
\begin{equation}
\label{formationbiastake2}
b_1(S_0, \zform) = 1 + 
{D(S_1, \zform) \over D(S_0, \zform)}
\left( {{\delta^2_{1, {\rm form}}}\over S_{1, {\rm form}}}
- 1 \right)
{1 \over \delta_{1, {\rm form}}} \, ,
\end{equation}
where we have assumed that at a sufficiently early $z_i$,
the growth factor is scale-independent, i.e. $D(S_0, z_i) = D(S_1, z_i)$.
This is a useful expression because
$\delta_{1, {\rm form}}$ is typically a constant, with fairly
weak dependence on $\zform$ and cosmology (see footnote
\ref{1686}). This appears to be a good approximation even
for the modified gravity model we will study below, where for sufficiently
small haloes $\delta_{1, {\rm form}}$ approaches its $\Lambda$CDM value,
which is $1.671$ for $\zform = 0$.

\subsection{The modified linear growth factor}

So far, we have not been very explicit about
where the scale-dependent growth factor actually comes
from. On the sub-Hubble scales of interest, mass and momentum
conservation imply the following
equation for the linear matter overdensity $\delta$:
\begin{eqnarray}
\frac{\dr^2 \delta}{\dr a^2} + \left(\frac{3}{a} + \frac{1}{H}\frac{\dr H}{\dr a}\right)
\frac{\dr\delta}{\dr a} = \frac{1}{a^4 H^2} \nabla^2\phi\,,
\label{eq:lingrowth}
\end{eqnarray}
where the Laplacian is in comoving coordinates.
What modified gravity, or clustering in dark energy,
does is to modify the relation between $\delta$
and the gravitational
potential $\phi$, from the standard Poisson equation.
The result can often be modelled as replacing
$\nabla^2 \phi$ by ${\cal O} \cdot \delta$,
where ${\cal O}$ is some linear operator which
can be scale-dependent, giving rise to a scale-dependent
growth for $\delta$.
Solving the resulting equation in Fourier space
would give $\delta(k, z) = 
[\wtilde D (k,z)/\wtilde D (k,z_i)] \delta(k,z_i)$, 
where $k$ is the wavenumber of interest. 
We use the symbol $\wtilde D$ to distinguish this
growth factor from 
the growth factor $D$ we use in the excursion set 
calculation. The difference arises from the fact
that the excursion set calculation cares about
the growth of the variance $S$, smoothed on scale
$R$, or equivalently, $M$. We define $D$ in such a way
to give the correct evolution of the variance $S$:
\begin{subequations}
\label{eq:Deff}
\begin{align}
&D^2 (R, z) \equiv \frac{S(R,z)}{S(R,z_i)}\,,\\ 
&\qquad = \int_0^{\infty} {\dr^3 k \over (2\pi)^3} P(k, z_i) 
\left\{\wtilde{W}(kR) \frac{\wtilde{D}(k, z)}{\wtilde{D}(k, z_i)}\right\}^2 \biggr/ \nonumber\\* 
&\qquad \qquad \qquad \int_0^{\infty} {\dr^3 k \over (2\pi)^3} P(k,z_i) \wtilde{W}^2(kR)\,.
\end{align}
\end{subequations}
In the excursion set computation, the overdensity $\delta$ is evolved
using the same growth factor $D$. In practice, 
the integral over $k$ is expected to be dominated by $k \sim 1/R$,
and therefore $D(R,z)$ is roughly $\tilde D(k=1/R, z)$, though
we do not use this approximation.

\subsection{\label{sec:calcf}Calculating the first-crossing distributions}

As mentioned above, the flat barrier approximation turns out
to be a fairly good one for the example we study. 
However, in general, the barrier or threshold for collapse is
mass/scale-dependent. The corresponding 
first-crossing distribution
can be found as follows.

First of all, it is possible to calculate the $f$ distribution analytically for barriers which are linear
functions of $S$: if $\delta_1(S) = \omega - \beta S$, 
\[
f(S_1,\delta_1|S_0,\delta_0) = \frac{ \Delta\delta}{\Delta S\sqrt{2\pi \Delta S}} \exp{\left\{ 
     -\frac{\left( \Delta\delta - \beta\Delta S \right)^2}{2\Delta S}\right\} }\,,
\]
where $\Delta\delta \equiv \delta_1(S_0) - \delta_0$ and $\Delta S
\equiv S_1 - S_0$ \cite{Sheth:1998ew}. 
For a more
general barrier, there is no exact analytic solution, but
reasonable approximations have been worked out
by Sheth \& Tormen, and Lam \& Sheth
\cite{Sheth2002,Lam2009}. 
In this paper, we adopt the algorithm of 
Zhang \& Hui \cite{Zhang:2005ar} which gives an
exact, albeit numerical, solution
for the unconditional $f$-function. It is straightforward to extend
their method to the conditional case. 
It can be shown that the function is the solution 
of the integral equation
\begin{align*}
f(S_1,\delta_1(S_1)|S_0,\delta_0) &= g_1[S_1,\delta_1(S_1);S_0,\delta_0] \\*
&\quad + \int_{S_0}^{S_1} \dr S'_1 \, g_2[S_1,\delta_1(S_1);S'_1,\delta_1(S'_1)]\\* 
& \qquad \times f(S'_1,\delta_1(S'_1)|S_0,\delta_0)\,,
\end{align*}
where
\begin{align*}
g_1[S_1,\delta_1(S_1);S_0,\delta_0] &= \left\{ \frac{\delta_1(S_1) - \delta_0}{S_1 - S_0} - 
2\frac{\dr \delta_1}{\dr S_1}\right\} \\* 
&\quad \times P_G(\delta_1(S_1)-\delta_0, S_1 - S_0)\,, \\
g_2[S_1,\delta_1(S_1);S'_1,\delta_1(S'_1)] &= \left\{2\frac{\dr \delta_1}{\dr S_1} - 
\frac{\delta_1(S_1) - \delta_1(S'_1)}{S_1 - S'_1}\right\} \\*
 & \quad \times P_G(\delta_1(S_1)-\delta_1(S'_1),S_1 - S'_1)\,.
\end{align*}

This is a Volterra equation of the second kind. By treating $f$ as a 
vector $\bb f$, we can see
that this equation is of the form 
$\bb f = \bb g + \bb{M f}$ where $\bb g$ is another vector
and $\bb M$ is a matrix, implying that $\bb f$ can
be found with a matrix inversion: $\bb f = (\bb I -\bb M )^{-1}\bb g$. 
$\bb M$ is a triangular matrix, allowing $\bb I - \bb M$ to be inverted efficiently 
by iteration \cite{numrecipes1992}.

\section{\label{sec:illustration} An illustration with a Yukawa model}

By the Yukawa model, we mean using the following
effective Poisson equation to compute the growth of perturbations
and the formation of haloes (for instance, it goes into the right
hand side of Eq. [\ref{eq:lingrowth}]):
\begin{eqnarray}
\label{yukawa}
-k^2 \phi = 4\pi\frac{G_N}{1+\alpha}\bar{\rho} a^2 \delta
\left[ 1 + \alpha\frac{k^2}{k^2 + a^2/\lambda^2 } \right]\, ,
\end{eqnarray}
which is written in Fourier space.
Here, $G_N$ is the effective Newton's constant at high $k$.
The $k^2/(k^2 + a^2/\lambda^2)$ term corresponds to a Yukawa modification,
where $1/\lambda$ can be thought of as a mass: it determines
on what scale modifications to GR become important.
The parameter $\alpha$ quantifies
the size of the modification, i.e. the ratio of effective
Newton's constant on small scales to on large scales is
$1/(1 + \alpha)$. A positive $\alpha$ corresponds to weakening
gravity on large scales (large compared to $\lambda$), 
while a negative $\alpha$ (but $> -1$) corresponds to the opposite.

This effective Poisson equation can be solved by
\begin{equation}
\phi(\bb r) = - \frac{G_N}{1+\alpha} \int \dr^3\bb r'\,
\frac{\bar\rho \delta(\bb r')}{|\bb r - \bb r'|}
\left\{1 + \alpha \exp{\left( - |\bb r - \bb r'|/\lambda\right)}\right\}\,, 
\label{eq:potential}
\end{equation}
where $\bb r$ is the physical proper coordinate, related to the
comoving coordinate $\bb x$ by $\bb r = a \bb x$.

The model encapsulated by Eq. (\ref{yukawa}) is
phenomenological in nature.
For a positive $\alpha$, it can be derived from a scalar-tensor
theory with a potential for the scalar field, described in
Appendix \ref{sec:scalartensor}.

One implication of Eq.~(\ref{eq:potential}) is that Birkoff's theorem, or Gauss's 
law, no longer holds: a test particle in a spherically-symmetric matter distribution 
feels a force from the matter both inside and outside the surface of constant radius 
defined by its position. 

We investigate three methods, one exact and two approximate,
for solving for the nonlinear evolution of 
spherically-symmetric perturbations,
the details of which are deferred to Appendix~\ref{sec:SC}. 
This nonlinear problem has to be solved to determine the appropriate
collapse threshold $\delta_{{\rm crit}, i}$ used in the excursion set calculation.
The violation of Birkoff's theorem implies an initially top-hat
perturbation does not remain so.
However, we find that modelling 
the spherical perturbation as homogeneous (i.e. a top hat)
at every time step is a good
approximation for the parameters and halo masses of interest.
Other infra-red modifications of gravity, such as DGP,
appear to share the same feature
---Schaefer 
\& Koyama have argued \cite{Schaefer:2007nf}
that the effect of the violation of Birkoff's theorem in DGP 
gravity should be small.

We find that a Yukawa-type modification of gravity, which is consistent with observations of galaxy
clusters and hence has $\lambda > 5 \hmpc$, will only affect the
collapse dynamics for haloes with mass $M \agt 10^{14} \hmsun$; for smaller haloes the evolution
is basically GR-like. This is consistent with previous studies of spherical collapse of large 
haloes in Yukawa-modified gravity \cite{Martino:2009}.

In the rest of this section, we present results for the barrier 
$\delta_{{\rm crit}, i}$ from
the spherical collapse computation, and for the halo bias $b_1$ from
the excursion set method. 
The background cosmology is taken to be that of $\Lambda$CDM
(see Appendix \ref{sec:scalartensor} for discussions on consistency
with modifying gravity for the perturbations). 
For the power spectrum, we use the analytic fit from \cite{Eisenstein:1997ik},
with parameters fixed by the WMAP 5-year analysis: $\Omega_{\rm CDM} = 0.206$,
$\Omega_{\rm baryon} = 0.043$, $\Omega_{\Lambda} = 1 - (\Omega_{\rm CDM} + \Omega_{\rm baryon})$,
$n_s = 0.961$, $h = 0.724$ \cite{Komatsu:2008hk}.
We normalize the power spectrum at a high redshift ($z_i = 100$), i.e. the
primordial normalization is the same for both the standard
and non-standard models. The amplitude is chosen such that,
if GR were to hold, $\sigma_8$ would equal $0.787$ today.

\subsection{\label{sec:barriers} Barriers}

\begin{figure}[tb!]
\subfigure[Examples for $\alpha = -0.5$, for different $\lambda$, as given in 
units of $\hmpc$ beside each curve.]{\includegraphics[width=8.6cm]{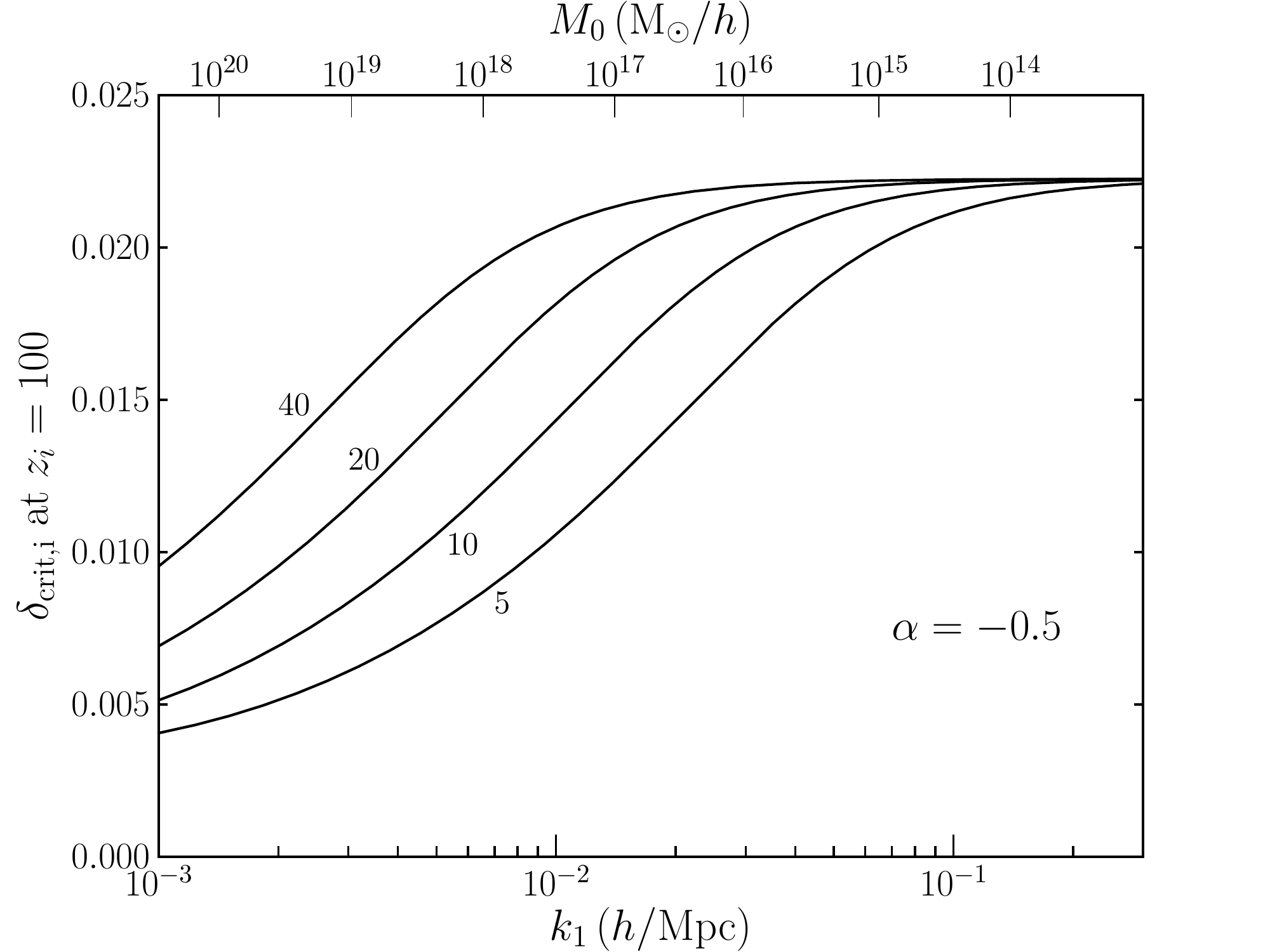}}
\subfigure[\label{fig:barrierlambda}Examples for $\lambda = 10 \hmpc$, for different $\alpha$, given beside each 
curve. Note the log scale on the y-axis.]{\includegraphics[width=8.6cm]{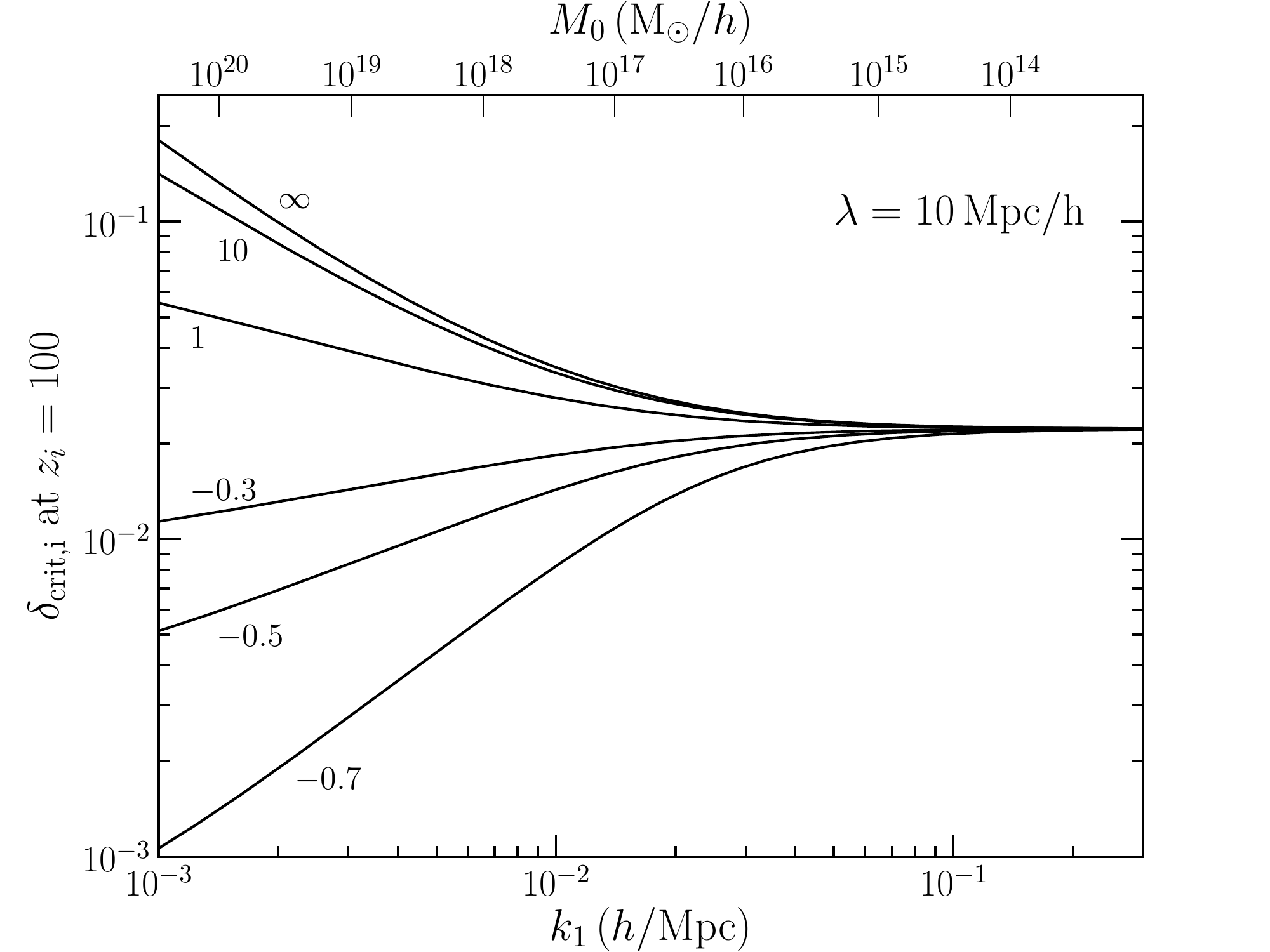} }
\caption{\label{fig:barrier} Barriers in Yukawa models
parametrized by $\alpha$ and $\lambda$:
the overdensity at $z_i = 100$
required on each comoving scale $k_1$ or mass
scale $M_1$, in order for that region to collapse by 
$\zform = 0$. }
\end{figure}

The barrier used in the random walk calculation is the initial overdensity required
for a perturbation to collapse to zero radius by an identification redshift $\zform$. In 
general relativity the required initial overdensity 
is independent of scale, i.e. the barrier is flat.
Curved barriers have been used to model the effect
of ellipsoidal, rather than spherical, collapse; e.g. see \cite{Sheth:1999su}.

For the Yukawa model, the barrier can be generated using the spherical collapse calculation
outlined in Appendix~\ref{sec:SC}; in what follows we use the homogeneous sphere 
approximation throughout. 
For each scale $k_1$ ($\equiv 1/R_1$, or a mass of $M_1$), 
the initial overdensity is found which collapses at $\zform$.
This defines $\deltac$, which is equated with
$\delta_1$ in the excursion set theory, as a function of $k_1$ or $M_1$.
FIG.~\ref{fig:barrier} 
shows barriers for the Yukawa model with a range of
parameter values for $\alpha$ and $\lambda$.
At smaller $k_1$ in the weaker-gravity case ($\alpha > 0$), the barrier
is higher because these larger-scale perturbations enter 
the weakened-gravity regime
earlier and experience more of a reduction in the gravitational force,
and so need a larger initial overdensity to push them over the edge 
into collapse; the converse holds for the stronger-gravity ($\alpha < 0$) 
model.
At high $k_1$ the barrier approaches the GR result, which can be thought
of as the value $1.671$, scaled back to
$z_i = 100$ using the $\Lambda$CDM growth factor ratio
$D(z_i)/D(\zform)$. 
While not a focus of this paper,
FIG. \ref{fig:barrier} suggests the mass function at the high
mass end would be significantly affected by the curving of the
barrier, especially its lowering at high mass for
$\alpha < 0$. The observed cluster count already constrains
$\alpha$ to be not too negative \cite{Martino:2009}.

\begin{figure}
\includegraphics[width=8.6cm]{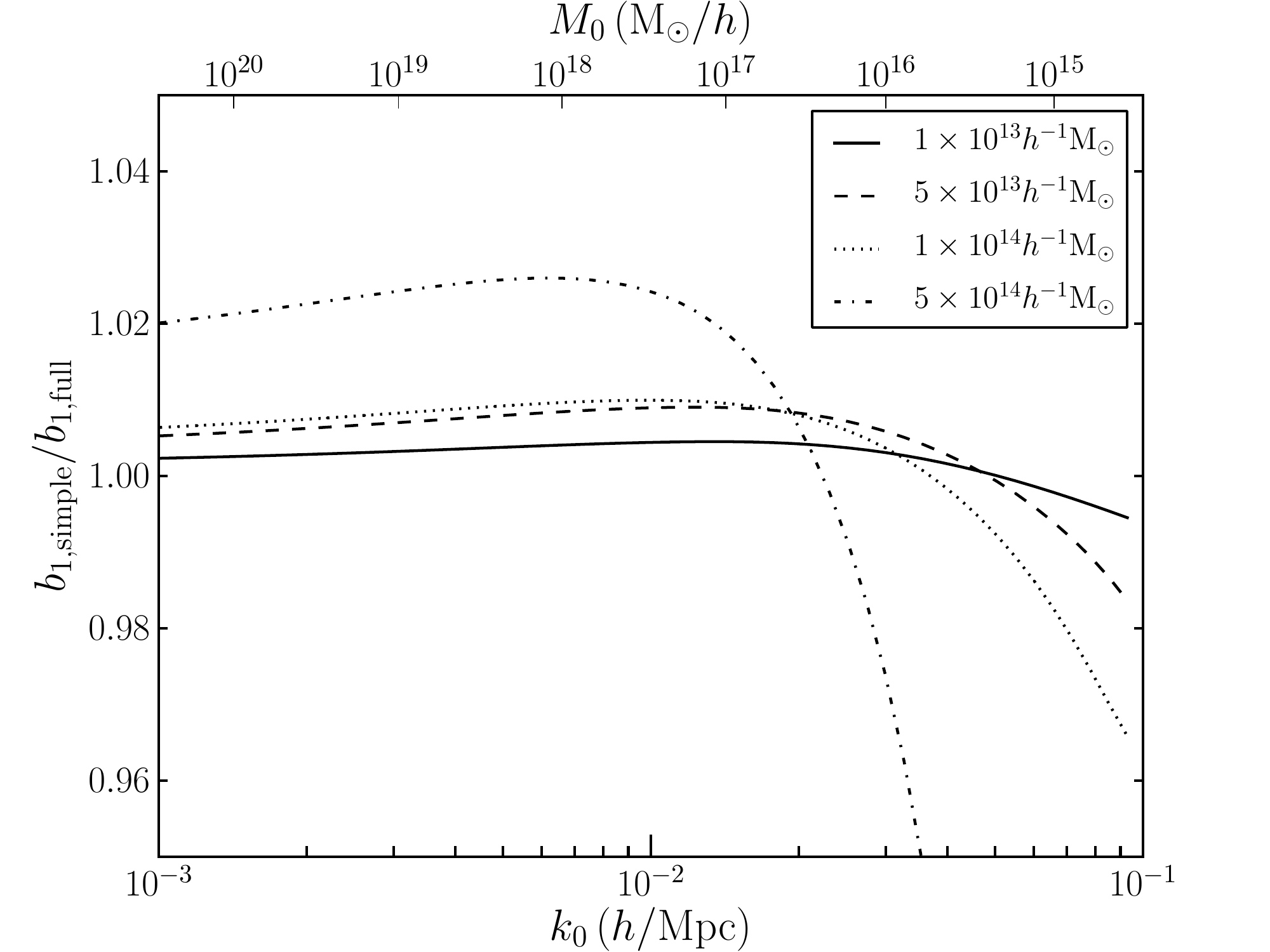}
\caption{\label{fig:bratio} Ratio of simple (flat barrier, Eq.~\ref{eq:bsimplefull}) to full
(curved barrier, Eq.~\ref{eq:b}) bias expressions, for $\alpha=-0.5$, $\lambda = 8 \hmpc$, 
$\zobs=\zform=0$.}
\end{figure}

Anticipating the linear halo bias that we will calculate next,
in FIG.~\ref{fig:bratio} we plot the ratio of the flat barrier solution, 
Eq.~(\ref{eq:bsimplefull}), to the full curved barrier solution, Eq.~(\ref{eq:b}) using the
numerical procedure of Sec.~\ref{sec:calcf}, for four halo masses. The amplitude of the flat 
barrier is taken to be the collapse threshold $\delta_1$ at high $k_1$, which in the case of
our models is very close to the Newtonian value.
The flat barrier approximation is
very good for galaxy-mass haloes---for a halo of $10^{13}\hmsun$, 
the fractional discrepancy in the bias is of order $10^{-3}$ for the parameters considered.
Decreasing $|\alpha|$, increasing $\lambda$, or increasing $\zform$ will flatten the barrier,
 further reducing this difference. Effects due to barrier curvature are more
readily apparent for cluster-mass haloes, because the barriers in FIG.~\ref{fig:barrier} are
only curved at large masses.

\subsection{\label{sec:birth} Formation bias}
There are two ways in which the halo formation bias (i.e.
halo bias with $\zform = \zobs$) differs from the
standard expression, due to for instance a modification to
the gravitational law.
One is from the scale-dependent linear growth, the other
is from the curved barrier. The former is the dominant effect.
As we have seen already from FIG.~\ref{fig:bratio}, the curvature
of the barrier does not have a significant impact on the halo bias,
on scales $k_0 \lsim 0.05$ h/Mpc. The impact on smaller scales
is higher, but still minor for sufficiently light haloes.

Our most general expression for the halo bias is
given by Eq. (\ref{eq:b}). 
The factor of $D(S_0, z_i)/D(S_0, \zobs)$ encapsulates the first
effect. 
For instance, in the Yukawa model with $\alpha > 0$ (weaker
gravity on larger scales), this factor is larger on large scales,
and the halo bias is correspondingly higher.
Physically, what drives the higher halo bias on
larger scales is this: recall that we are interested in
how the halo overdensity $\delta_h$ relates to the
matter overdensity $\delta$ at redshift $\zobs$;
for a fixed $\delta$, the overdensity $\delta_0$ at the
initial redshift $z_i$ (where the random walk is performed)
is higher on larger scales if $\alpha > 0$.
This larger $\delta_0$ boosts halo formation, and results
in a higher halo bias.

The curved barrier is in principle another way in which
a scale dependence is imprinted on the halo bias.
For this to occur, the barrier must absorb some trajectories before they reach ($S_1$,$\,\delta_1(S_1)$) in a way
which depends on the starting point $S_0$. We do not observe this in the models we
consider, because on large scales the random walk has a negligible probability of
crossing the barrier because the variance of each step, $S$, is so low. Therefore 
large-scale changes to the barrier have very little effect on the bias. (If $\alpha$ is
strongly negative, say $\alpha\sim -0.7$, the barrier is so low on large scales that it 
absorbs the random walk paths before they reach `sensible' halo scales, see
FIG.~\ref{fig:barrierlambda}. This produces a strongly scale-dependent bias and
drastic under-production of low-mass haloes, which is clearly incompatible with observations.)

\begin{figure}
\subfigure[\label{fig:birthpos} $\alpha = 1.0$, $\lambda = 10 \hmpc$.]
{\includegraphics[width=8.6cm]{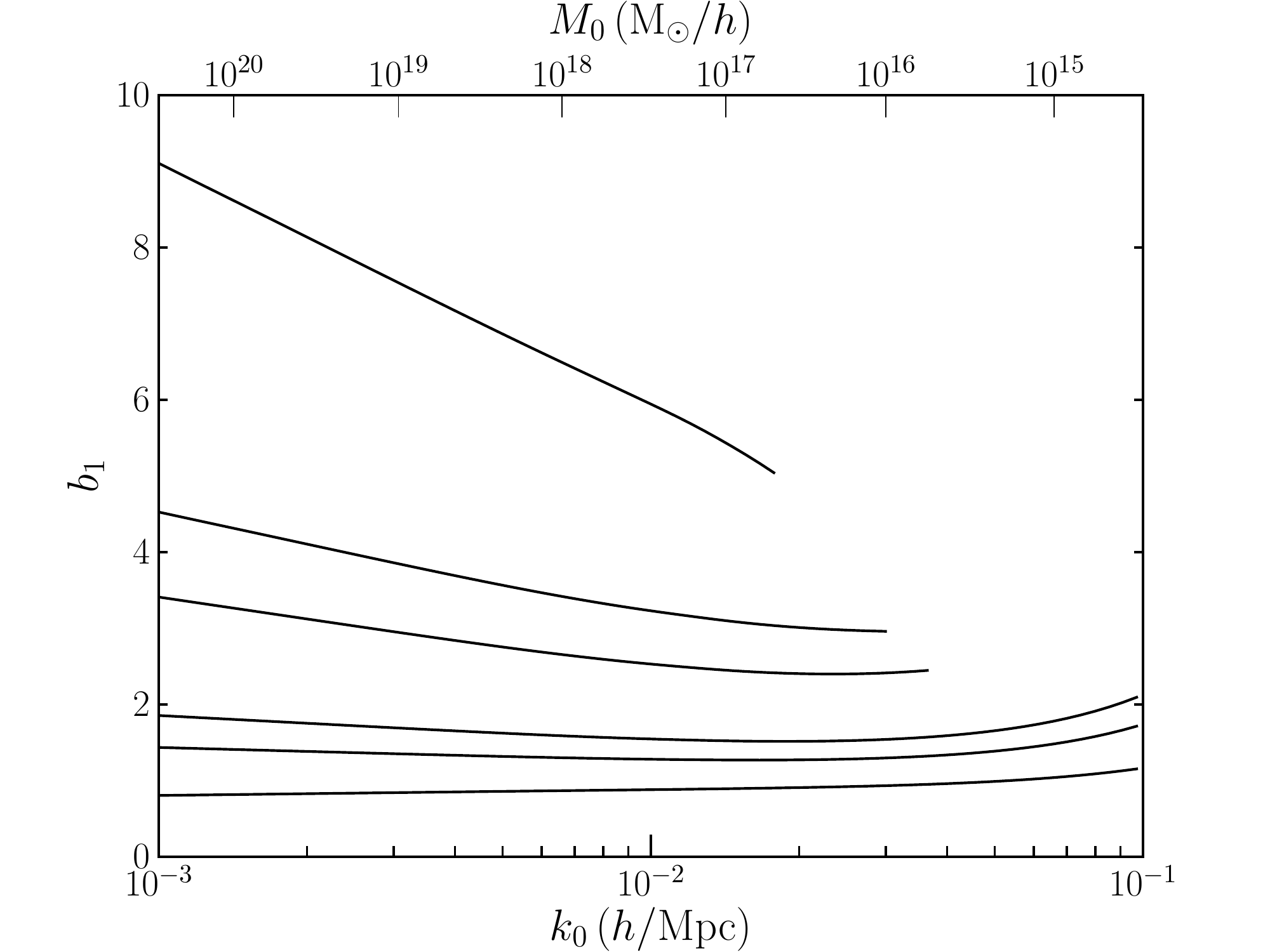}}
\subfigure[\label{fig:birthneg} $\alpha = -0.5$, $\lambda = 10 \hmpc$.]
{\includegraphics[width=8.6cm]{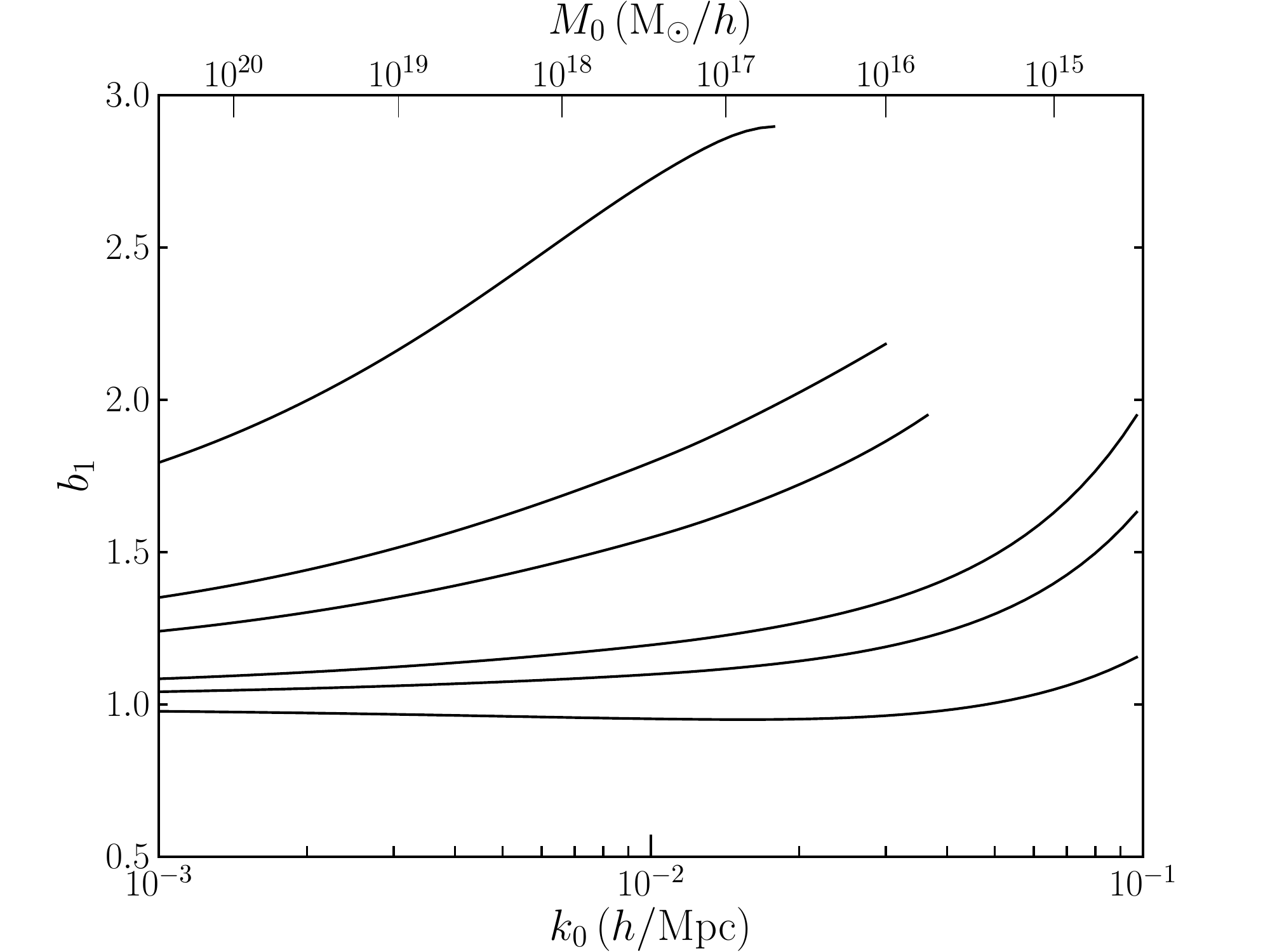}}
\caption{\label{fig:birthscale} Bias versus scale for $\zform = \zobs = 0$. 
The halo masses in each case are, in units of $\hmsun$, $5\times 10^{14}$ 
(most biased), $10^{14}$, $5\times 10^{13}$, $10^{13}$,
 $5\times 10^{12}$, and $10^{12}$ (least biased). For the three largest masses,
the curves are discontinued beyond the scale at which halo exclusion 
becomes important, i.e. where $M_1/M_0 = 0.01$.
The panels illustrate the predictions for two Yukawa models.
}
\end{figure}

FIG.~\ref{fig:birthscale} shows the halo bias 
computed using the full excursion set theory, with a barrier
determined by the exact spherical collapse calculation.
The scale dependence of the bias
comes almost entirely from the large-scale growth factor, and 
is well described by
our flat-barrier expressions, Eqs.~ (\ref{eq:bsimple}), (\ref{formationbias}), or (\ref{formationbiastake2}). 
As the halo mass increases (and so the
variance $S_1$ decreases) the factor multiplying the scale-dependent growth factor ratio,
$({\delta_1}^2/S_1 - 1)/\delta_1$, becomes larger, leading to more pronounced scale dependence in
the formation bias. The different curves in FIG.~\ref{fig:birthscale} are produced by
the same scale-dependent factor multiplying a mass-dependent amplitude.

\begin{figure}
\includegraphics[width=8.6cm]{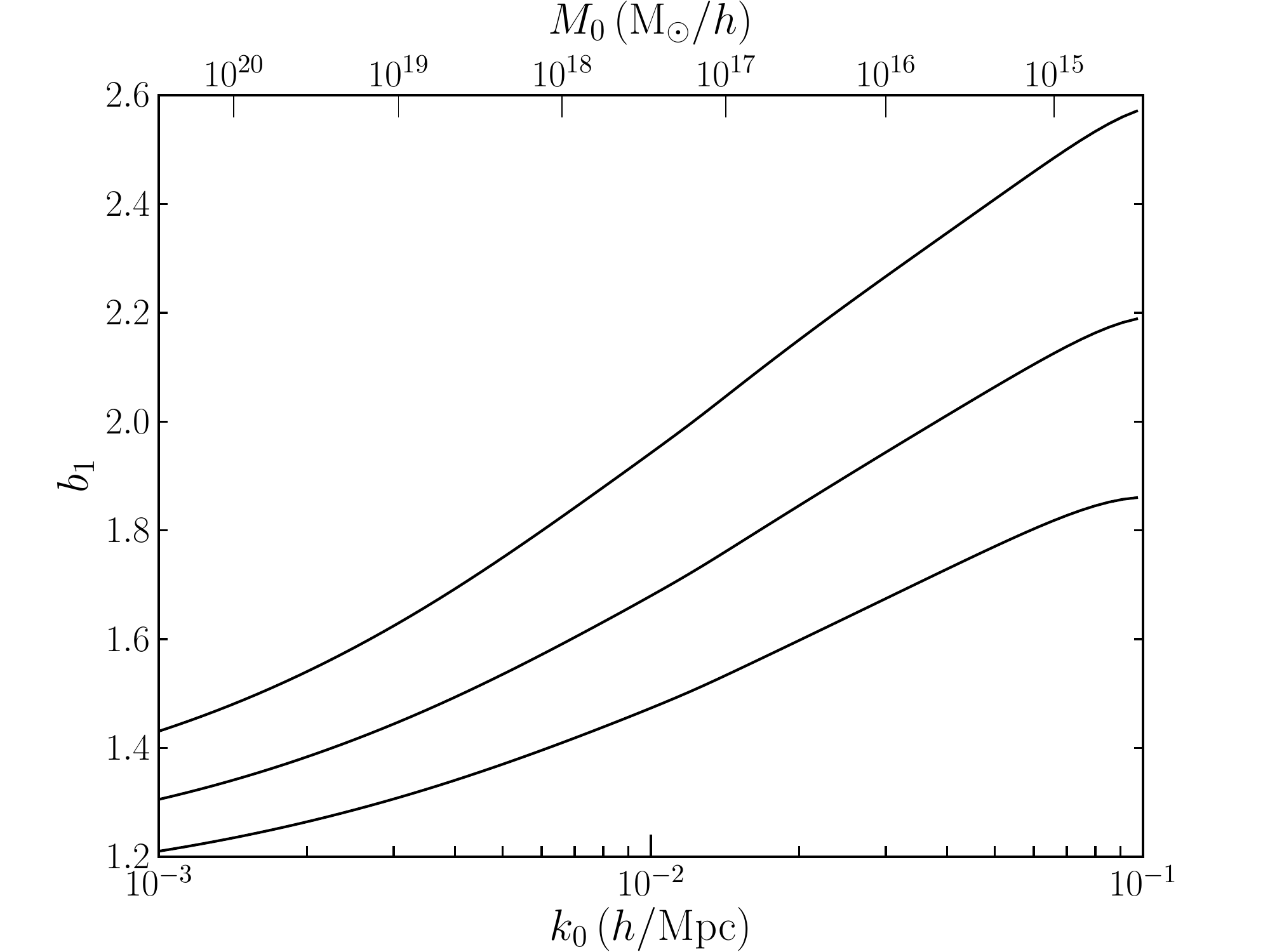}
\caption{\label{fig:passive} Passive bias evolution of a $5\times 10^{12}\hmsun$ halo which forms
at $\zform = 1$ and is observed at $\zobs = 1$ (top), 0.5, and 0 (bottom); in this case
$\alpha = -0.5$, $\lambda = 10 \hmpc$.}
\end{figure}

\subsection{\label{sec:passive} Passive bias evolution}

In the previous section we observed the haloes at the time of their formation. Some galaxy
populations form at an early redshift and then evolve passively, without further 
major merger events. In this scenario the total number of haloes is conserved after the 
formation epoch; the haloes increase their mass only by accreting much smaller haloes. The
mass assigned to the haloes is their mass at formation, disregarding subsequent accretion.

Here $\zform \neq \zobs$; the formation redshift determines the barrier shape and 
height via $\delta_1 = \deltac$ (this is the initial overdensity threshold
which has to be evolved back from the formation redshift), while the 
observation redshift determines the relationship 
between $\delta_0$ and $\delta$, Eq.~(\ref{eq:delta}). This combination reproduces, in the
excursion set formalism, the modified-gravity bias-evolution results previously derived from
the perturbation equations by HP \cite{Hui:2007zh},
in the sense that $b_1 - 1$ scales with $D(S_0, \zform)/D(S_0, \zobs)$,
as in Eq. (\ref{eq:oldbias}). However, it should be kept in mind
that according to the excursion set theory, the formation bias is not
scale-independent in general, but is instead given by Eq.~(\ref{formationbias}). 

The combined effects of formation bias and passive bias evolution can be non-intuitive. For
example, in FIG.~\ref{fig:passive} the stronger large-scale gravity 
($\alpha < 0$) causes the bias to 
be smaller on large scales, but passive evolution decreases the smaller scale bias faster. 
This is because on smaller scales the bias is greater, and thus is more 
quickly diluted by passive evolution.

\subsection{\label{sec:lrg}Comparison with observations --- an example}

For comparison with observations, it is sometimes convenient to
consider an apparent bias $b_{\text{app}}$, defined as
\begin{eqnarray}
\label{eq:bapp}
b_{\text{app}}(k_0) \equiv \frac{D(k_0,\zobs)}{D(k_0,z_i)}\frac{D_{\Lambda{\rm CDM}}(z_i)}{D_{\Lambda{\rm CDM}}(\zobs)} b_1(k_0) \, ,
\end{eqnarray}
where $D_{\Lambda{\rm CDM}}$ is the growth factor according
to the standard $\Lambda$CDM model, and $D$ is that according
to our non-standard model. This is constructed such that the power spectrum 
of haloes in the non-standard model $P_{\rm haloes}$ is
related to the matter power spectrum in $\Lambda{\rm CDM}$ 
$P_{\Lambda{\rm CDM}}$ by
\begin{eqnarray}
P_{\text{haloes}} = b_{\text{app}}^2(k) P_{\Lambda\text{CDM}} \, .
\end{eqnarray}
This is a necessary step when we don't know the matter power spectrum directly, and so must compare
the observed galaxy distribution to what would be expected from an underlying standard-gravity matter
distribution. If the matter power spectrum were known, 
from weak lensing for example, we could compare the
measured bias to the true bias $b_1$ directly.

\begin{figure}
\subfigure[$\alpha = 1.0$, $\lambda = 10 \hmpc$. Compare with
FIG.~\ref{fig:birthpos}.]{\includegraphics[width=8.6cm]{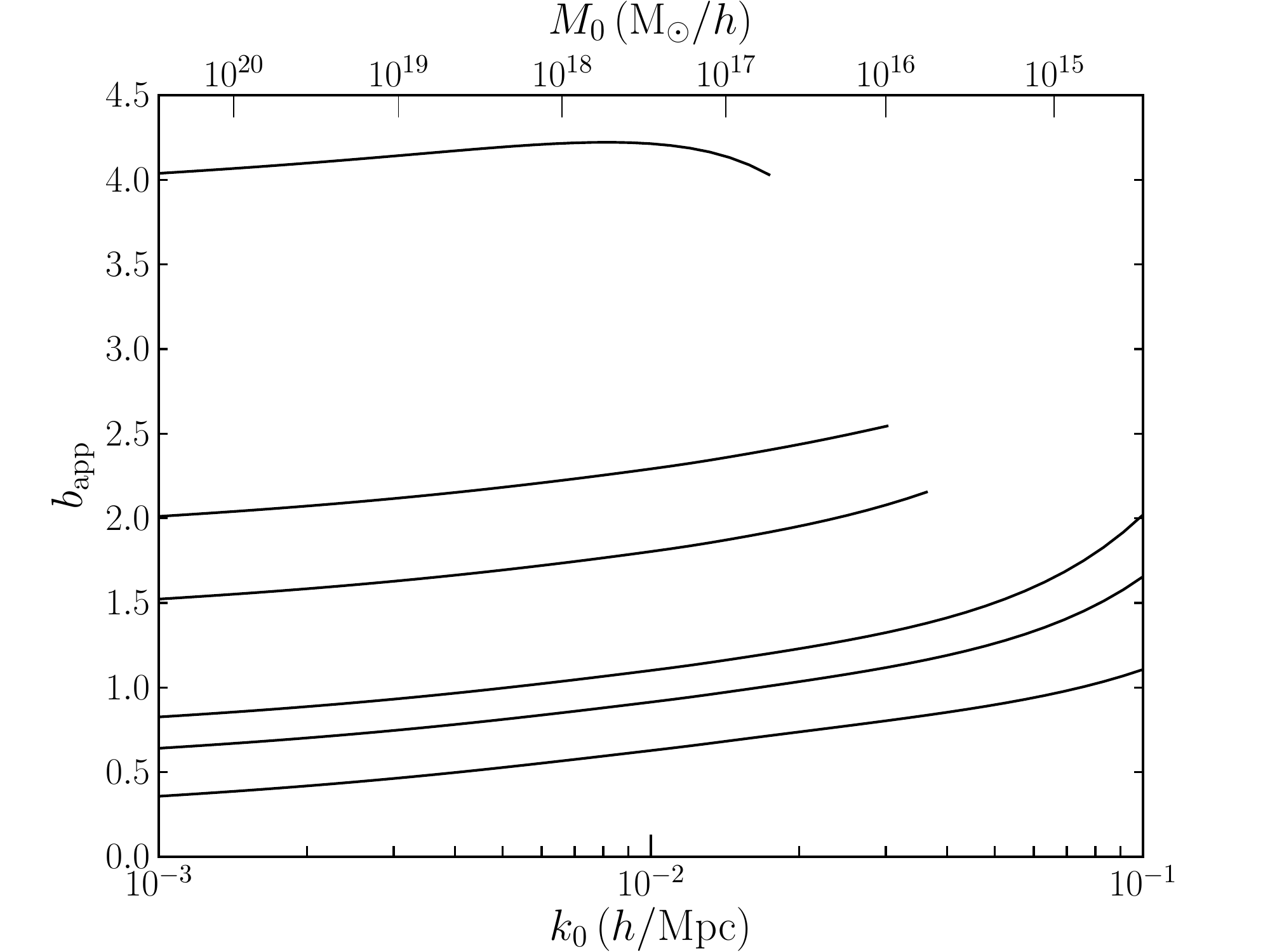}}
\subfigure[$\alpha = -0.5$, $\lambda = 10 \hmpc$. Compare with
FIG.~\ref{fig:birthneg}.]{\includegraphics[width=8.6cm]{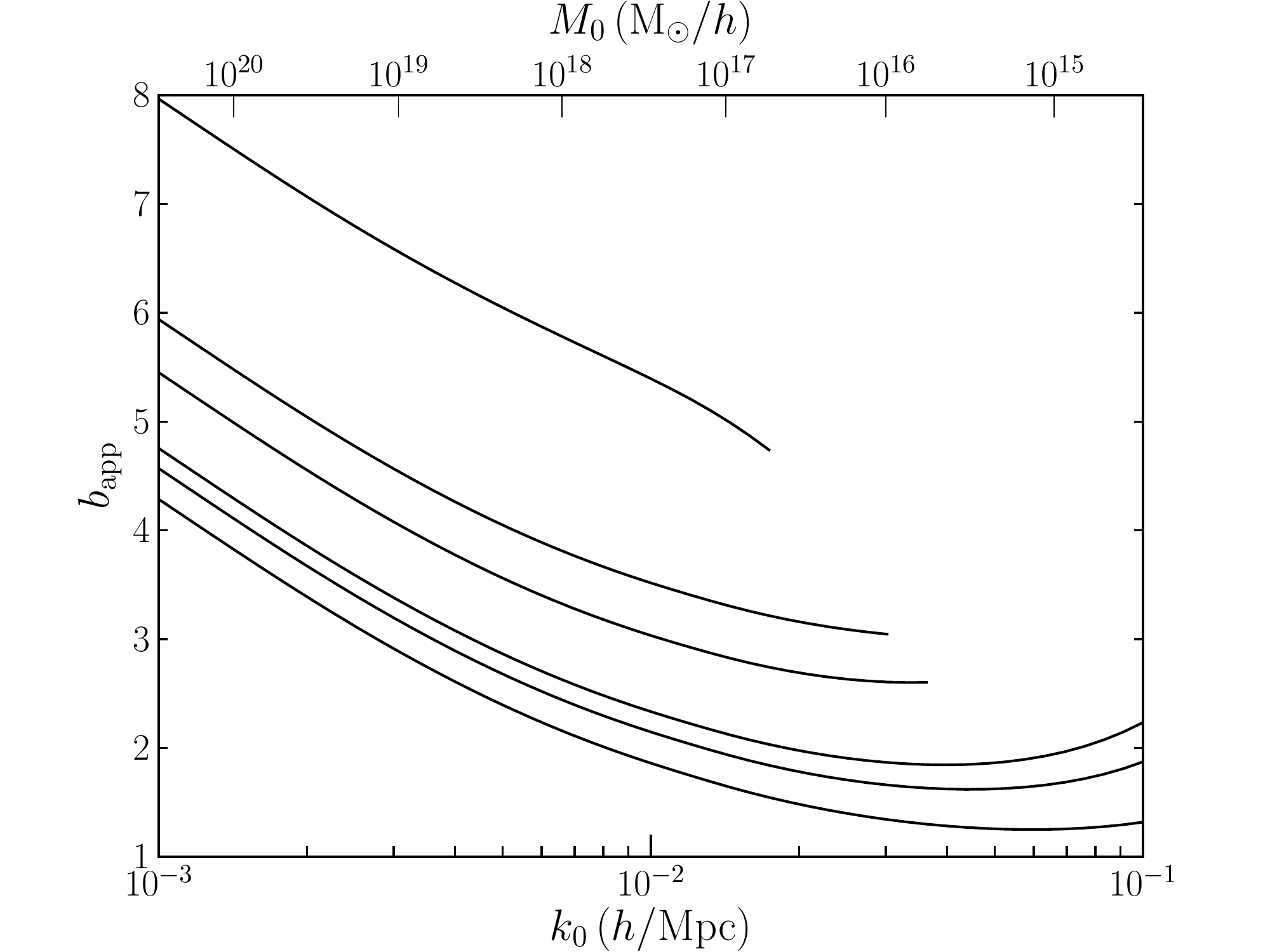}}
\subfigure[\label{fig:apppassive}Passive apparent bias evolution of a 
$5\times 10^{12}\hmsun$ halo;
$\zform = 1.0$, $\zobs = 1.0$ (solid), 0.5 (dashed), 0 (dashed-dotted). 
Compare with FIG.~\ref{fig:passive}.]
{\includegraphics[width=8.6cm]{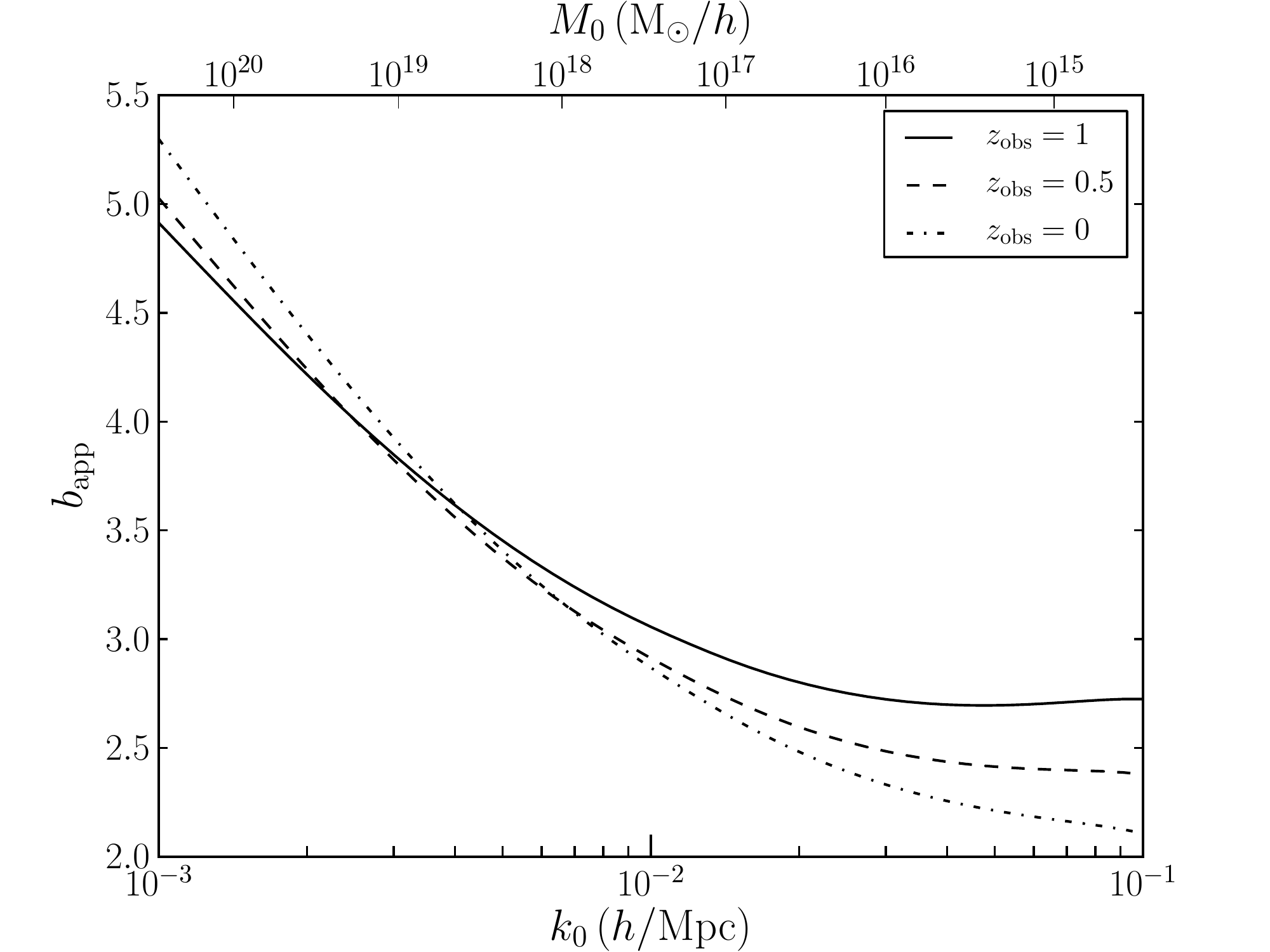}}
\caption{\label{fig:apparent} Birth and evolved apparent (rather 
than physical) biases for the same parameters
and halo masses as FIG.~\ref{fig:birthscale} and FIG.~\ref{fig:passive}.}
\end{figure}

The apparent bias is plotted against scale in FIG.~\ref{fig:apparent}. On large 
scales the apparent bias has a constant slope, which is the same for all halo
masses. This is because the form of the true bias, $b_1 \approx 1 + f_1(k_0)f_2(M_1)$ (Eq.~\ref{eq:bsimple}),
 when multiplied by the scale-dependent factor $D(k_0,\zobs)/D(k_0,z_i)$, becomes 
 $b_{\text{app}} \approx g_1(k_0) + g_2(M_1)$.

In FIG.~\ref{fig:apppassive}, the large-scale apparent bias actually \emph{increases}
with passive evolution---this is due to the overall multiplicative factor  
in front of $b_1$ in the previous equation, which is 
greater than one on large scales when $\alpha < 0$. The opposite occurs for positive $\alpha$:
the large-scale apparent bias drops faster than the physical bias.

\begin{figure}[tb!]
\includegraphics[width=8.6cm]{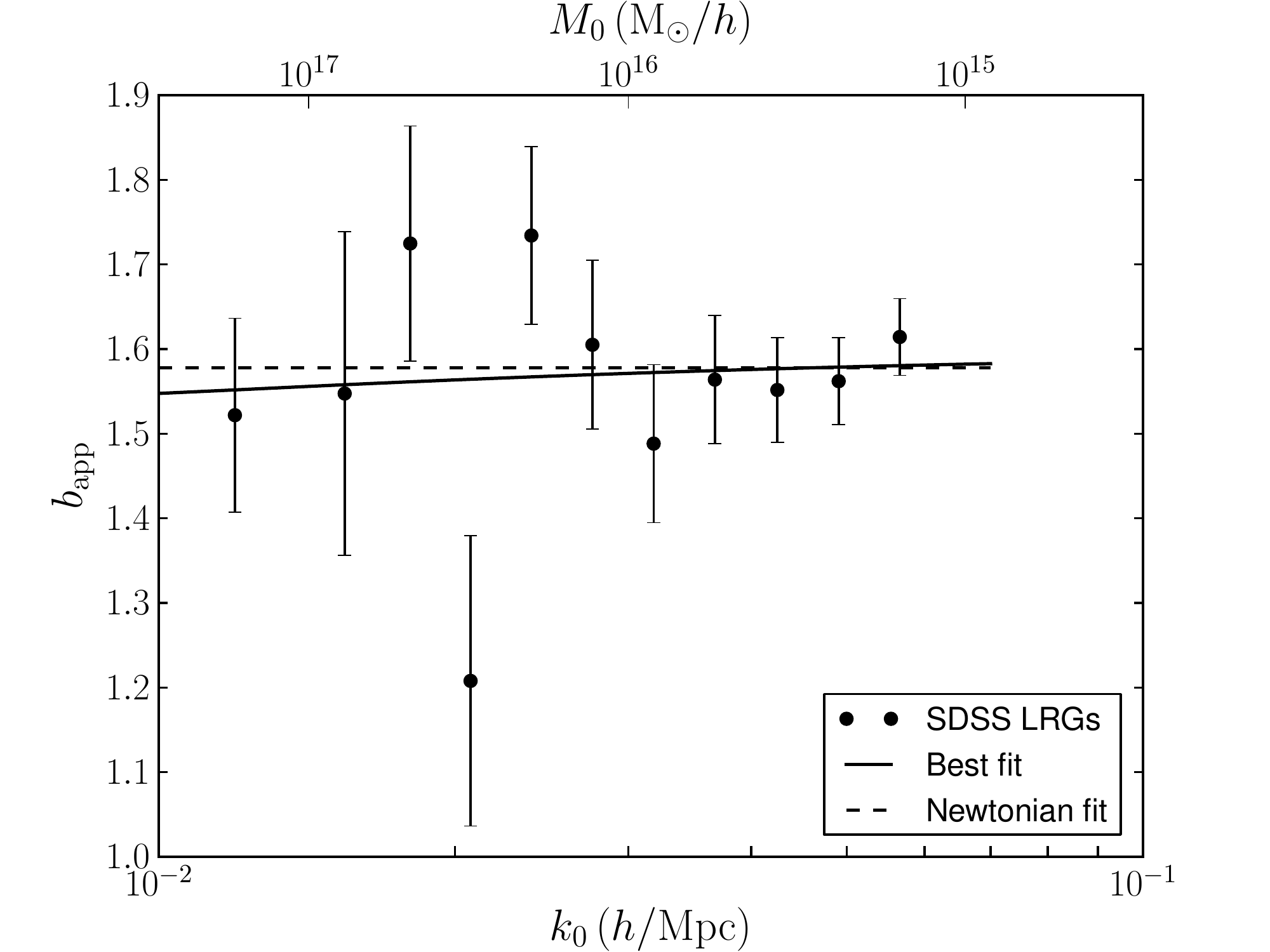}
\caption{\label{fig:bmeasure} Galaxy biases derived from the SDSS LRGs, with Yukawa
and Newtonian best fit curves. The best fit parameters are $\alpha=0.065$, 
$\lambda = 6.5 \hmpc$}
\end{figure}

\begin{figure}[tb!]
\subfigure[$\alpha > 0$]{\includegraphics[width=8.6cm]{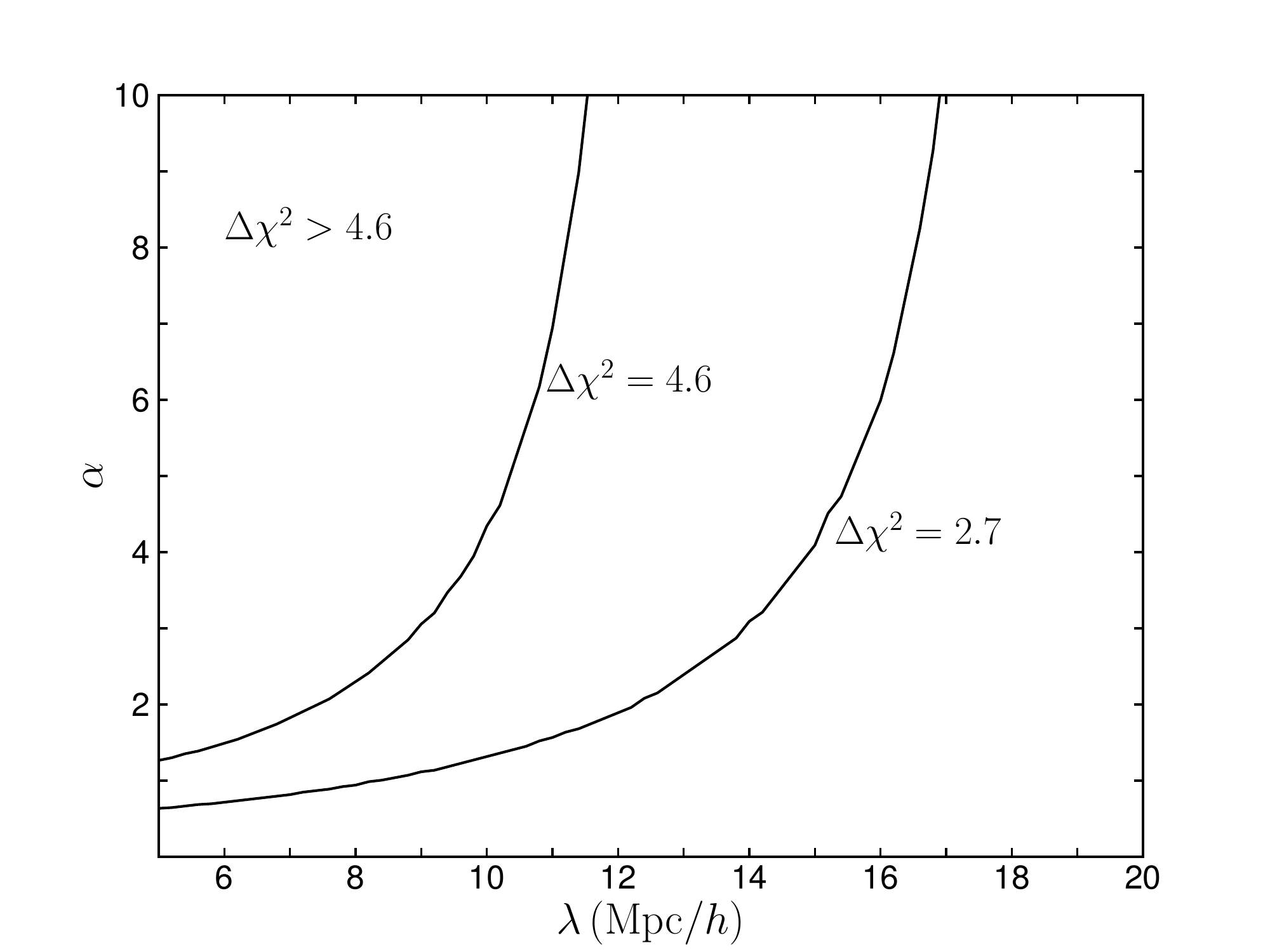}}
\subfigure[$\alpha < 0$]{\includegraphics[width=8.6cm]{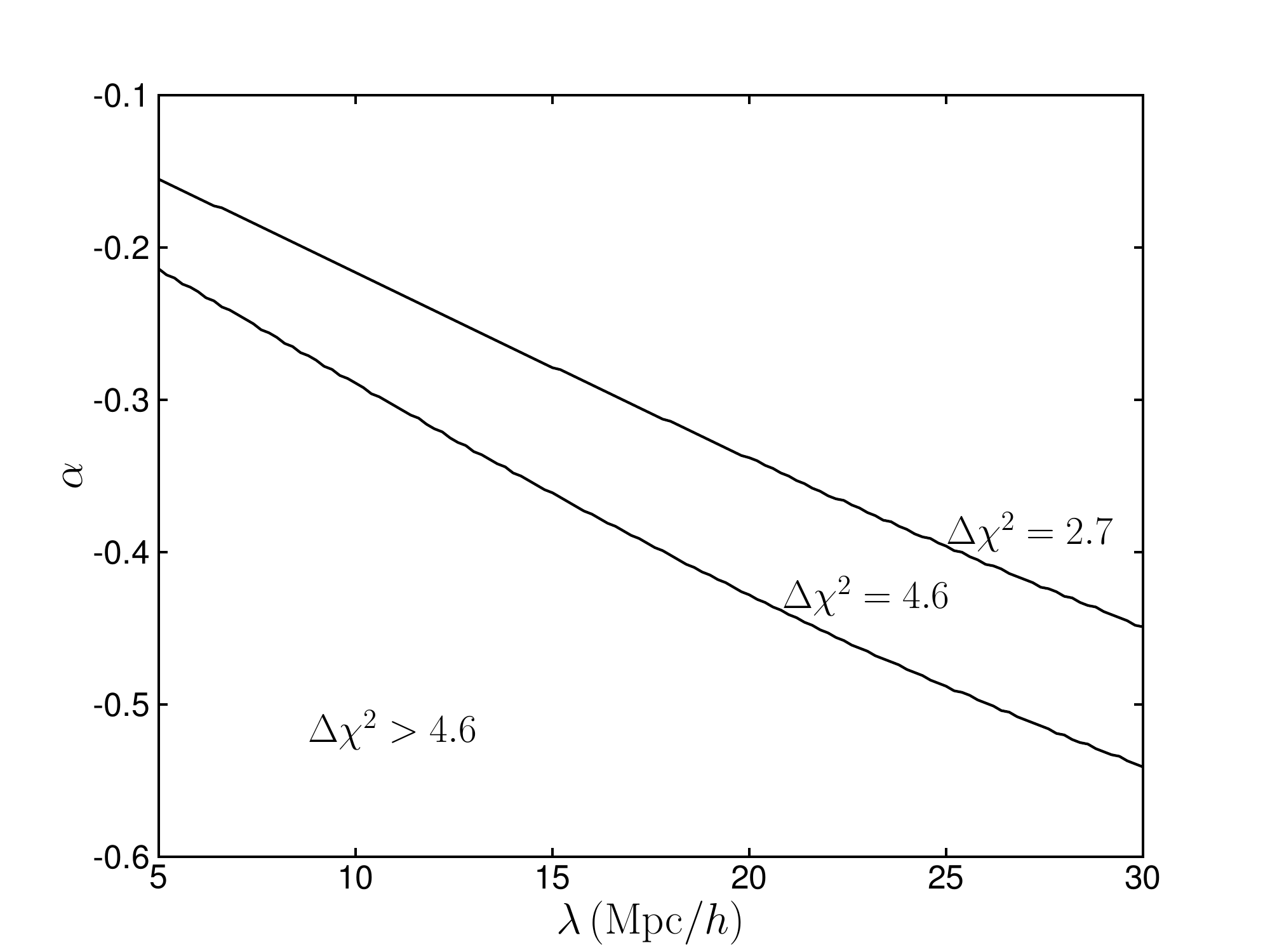}}
\caption{\label{fig:contours} Contours of constant $\Delta\chi^2$ in the
$\alpha$--$\lambda$ plane. The regions containing $\alpha=0$ are `included' by
the contours.}
\end{figure}

The high intrinsic luminosity of luminous red galaxies (LRGs) makes them useful
tracers of the large-scale cosmic density field. We calculate the galaxy bias of the 
LRGs as a function of scale, using the LRG power spectrum
derived from the SDSS galaxy redshift survey \cite{Tegmark:2006az}, and the standard 
matter power spectrum, whose parameters are set to the WMAP 5-year 
maximum-likelihood values \cite{Komatsu:2008hk}. The LRGs
are mostly believed to occur 
as the largest galaxy in individual haloes, 
and so the galaxy bias is approximately the host halo bias.  

A description of the halo clustering with only the first bias parameter and the 
matter power spectrum is consistent only on linear scales, which we estimate here
as those scales where the scale dependence introduced into the bias by a non-zero
$S_0(k_0)$ in Eq.~\ref{eq:bsimplefull} is small. This restricts our analysis of the LRG
power spectrum to the eleven data points available for $k_0 \lesssim 0.06 \hovermpc$.

The LRGs may be a passively evolving population \cite{Masjedi:2007cd} 
(however, see \cite{Wake:2008mf} for an
opposing view). Despite this, we use the formation bias at the median redshift of the 
sample, $\zform = \zobs = 0.35$, in order to include any mass the haloes may have accreted since
they formed. Ideally, one would bin the galaxies by redshift and account for the changing
predicted bias shape for each modified gravity model, thus providing two tests:
the shape of the bias curve at each redshift, and the evolution of the bias (in both
shape and magnitude) with time. This may be feasible with future data sets. 
 
A complete analysis would include marginalization over the matter spectral index 
and $\sigma_8$. To derive some preliminary constraints, 
we fix all the properties of the 
matter power spectrum and ($\Lambda$CDM) expansion history, and marginalize 
over halo mass in a $\chi^2$ minimization procedure\footnote{Here, we do not distinguish between the $k$'s
in our equations (which are really defined in real space, in the sense
of $k = 1/R$) and the true Fourier-space $k$'s as in the observed galaxy power
spectrum.  This is not an issue if one has a scale-independent growth factor, and therefore a
scale-independent large-scale bias.
But for our present purpose, some extra care should in principle be taken in
determining the correct mapping between our $k$'s and the observed
Fourier-based $k$'s. Simulations can be used to address this issue,
which we leave for future work.}  for each combination of 
$\alpha$ and $\lambda$.
Fitting the apparent formation bias at the median redshift of
the LRG sample gives best-fit parameters $\alpha = 0.065$, 
$\lambda = 6.5 \hmpc$, $M_{1} = 1.2\times 10^{13} \hmsun$, with $\chi^2 = 10.08$. 
This is statistically indistinguishable from the Newtonian fit, with $\chi^2 = 10.21$.
In fact, one could argue that the data does not favor the Yukawa model since
it requires more parameters without improving $\chi^2$ much.
However, the error bars are at the moment rather large.
The observed biases and model fits are shown in FIG.~\ref{fig:bmeasure}. 

FIG.~\ref{fig:contours} shows contours of constant $\Delta\chi^2$ in the $\alpha$--$\lambda$
plane. Marginalization over $\alpha$ or $\lambda$ is poorly constraining, since in both
cases the Newtonian limit ($\alpha\rightarrow 0$ or $\lambda\rightarrow\infty$) is 
allowed, and so the marginalized $\Delta\chi^2$ cannot be larger than 0.13. 
This method places better limits on stronger-than-Newtonian-gravity models.
These constraints should be considerably improved
by future galaxy clustering measurements, especially
those on scales $k_0 < 0.1 \hovermpc$.

\section{\label{sec:conclusions}Conclusions}

We have worked out the large-scale halo bias $b_1$
in a theory that has a
scale-dependent linear growth, such as modified gravity or clustered
dark energy. We give several expressions in \S \ref{sec:bias} \&
\ref{sec:flatbarrier}. 
The most general expression is Eq. (\ref{eq:b}). It relates $b_1$
to the first crossing distribution $f$, which can be computed using
the method described in \S \ref{sec:calcf}, for a general curved
barrier. Note that all quantities describing the random walk: 
$\delta_1, S_1, \delta_0, S_0$ are defined at the initial redshift
$z_i$, as opposed to the formation redshift $z_{\rm form}$. 

A much simpler expression, Eq. (\ref{formationbiastake2}), obtains
in cases where the barrier can be approximated as flat. As can be
seen in the example depicted in FIG. \ref{fig:barrier}, the barrier is
generally not flat once a scale-dependent growth factor is allowed.
However, it is often true that the barrier deviates from flatness only
for the largest masses, and a flat barrier approximation
(set at the small mass level)
actually works quite well in predicting the halo bias for more modest
halo masses (FIG. \ref{fig:bratio}). 
The expression in Eq. (\ref{formationbiastake2}) is particularly easy
to use because the threshold for collapse $\delta_{1, {\rm form}}$
and the variance $S_{1,{\rm form}}$ are defined at the formation
redshift as usual. 

Our main result is the imprinting of a 
scale dependence on the formation bias, by the modified linear growth 
factor. This can be understood to occur because regions of the same overdensity 
but different sizes today would not all have had the same overdensity
at an earlier redshift, and hence would have varying halo densities (see 
Sec.~\ref{sec:birth}). The scale dependence of the growth factor manifests 
itself at a single redshift, making it possible to look for it with a 
local-universe snapshot. Using this effect and an appropriate galaxy sample, 
one can place constraints on gravity theories, as we demonstrate with a simple 
Yukawa theory.

We have focused on the case in which the formation bias is deterministic.  However, 
this bias may be stochastic, in which case this stochasticity will also be scale-dependent \cite{Hui:2007zh}.  
Although we will not perform a detailed analysis of stochastic bias here, we note that, even if the formation bias is
 deterministic in, say, $k$-space, the scale dependence means that it will be stochastic in real space 
 (and vice-versa) \cite{Desjacques:2010}.  Thus, the stochasticity of the bias is another potential 
 signature of modified gravity theories. 

Looking forward, it would be interesting to work out how
a scale-dependent large-scale bias can be incorporated into
optimal weighting schemes for recovering the mass distribution
from a galaxy or halo catalog \cite{Hamaus2010,Cai2010}.

\acknowledgements

We would like to thank Zoltan Haiman for useful comments and encouragement.
LH thanks members of the CCPP at NYU, the IAS at Princeton
and Physics Department at the University of Hong Kong
for their fabulous hospitality.
This work is supported by the DOE DE-FG02-92-ER40699, NASA 09-ATP09-0049,  NSF AST-0908421,
and the Initiatives in Science and Engineering Program
at Columbia University.

\appendix
\section{\label{sec:extrap} Density field extrapolation}

In an excursion set calculation, there are in principle four different
redshifts that might be relevant: the initial redshift $z_i$ 
(taken to be $100$ in this paper), the formation redshift $z_{\rm
  form}$, the observation redshift $z_{\rm obs}$, and the redshift
at which all the random walk variables (the variances $S_0$, $S_1$, etc)
are defined: let us call it $\zfields$, or the random walk redshift.

In the standard theory, i.e. GR, the choice of $\zfields$ is
immaterial. The simplest way to see this is to examine the
flat barrier (but otherwise general)
expression for bias in Eqs. (\ref{eq:oldbias}) and
(\ref{formationbias}). In this expression,  the random walk
variables $\delta_1$ and $S_1$ are defined at the initial redshift
$z_i$. In other words, the choice $\zfields = z_i$ has been made.
It is simple to see that if the growth factor $D$ were scale-independent,
choosing a different random walk redshift, i.e. $z_i \rightarrow
\zfields \ne z_i$ would have made no difference to the predicted bias.
This is because in the expression for the formation bias
(Eq. [\ref{formationbias}]), $\delta_1^2/S_1$ is independent of the
random walk redshift by definition, and $\delta_1 D(S_0, z_{\rm
  form})/D(S_0, \zfields)$ (with $\delta_1$ defined at $\zfields$) is
also independent of $\zfields$. The key is that the growth factor
relevant for $\delta_1$, which is on scale $S_1$, is the same as
the growth factor on scale $S_0$. This is true in GR, but not true
for non-standard theories where the growth factor is generally
scale-dependent. Indeed, in the standard theory, it is common practice
to choose $\zfields = \zform$, such that for instance in an
Einstein-de Sitter universe, $\delta_1$ takes the value $1.686$.

In non-standard theories, the formation bias given by Eq.~(\ref{formationbias}) 
is still insensitive to the precise choice
of $\zfields = z_i$, as long as $z_i$ is sufficiently early. 
This is because we expect the (sub-Hubble) growth to be scale-independent
at sufficiently early times. Choosing $\zfields$ to be some late
redshift, however, would change the predicted bias. 
Our choice of $\zfields$ being equal to some early initial redshift is
in keeping with the excursion set philosophy, namely that
characteristics of the initial random Gaussian fluctuations, such as how
they change with smoothing scale (i.e. the random walk), determine
whether a region ultimately collapses into a halo by a certain redshift.

\section{\label{sec:scalartensor} Scalar-tensor Theory}

In this section, we would like to make explicit how the Yukawa model
we use as an illustrative example can be obtained from a scalar-tensor
theory \cite{Wagoner:1970vr}. In the Einstein frame, a scalar-tensor theory with
 potential scalar self-interactions
can be written as 
\begin{eqnarray}
&& S = {1\over 8\pi G} \int d^4 x \sqrt{-g} 
\left[ {1\over 2} R - {1\over 2} \nabla_\mu \varphi \nabla^\mu \varphi
- V(\varphi) \right] \nonumber \\ && + \int d^4 x {\cal L}_m (\psi_m,
\Omega^{-2}(\varphi) g_{\mu\nu}) \, ,
\end{eqnarray}
where $R$ is the Ricci scalar, the dimensionless
$\varphi$ mediates a scalar fifth
force, $V(\varphi)$ is the scalar potential, and
$\psi_m$ represents matter which is minimally coupled not
to the Einstein frame metric $g_{\mu\nu}$ but its conformal cousin
$\Omega^{-2}(\varphi) g_{\mu\nu}$. Here, $\varphi$ is typically small
such that the conformal factor can be expanded as
\begin{eqnarray}
\Omega^2 (\varphi) \sim 1 - 2 \beta\varphi
\end{eqnarray}
where $\beta$ is a coupling constant, typically of order unity for
a gravitational-strength scalar force. 

The total gravitational $+$ scalar acceleration an infinitesimal particle 
experiences is
\begin{eqnarray}
\ddot X_i = - \partial_i \phi = - \partial_i \tilde \phi- \beta \partial_i \varphi \, ,
\end{eqnarray}
where $\phi$ is the effective (total) gravitational potential, and
$\tilde\phi$ is the (Einstein frame) gravitational potential.
The scalar field $\varphi$ and the gravitational potential
$\tilde\phi$ satisfy the equations:
\begin{eqnarray}
&& \nabla^2 \tilde \phi = 4 \pi G \bar\rho a^2 \delta \nonumber \\
&& \nabla^2 \varphi = a^2\left[ {\partial V \over \partial \varphi}
- {\partial V \over \partial \bar\varphi}\right] + 8\pi \beta G \bar\rho a^2 \delta
\end{eqnarray}
where $\delta$ is the overdensity, $\bar\rho$ is the mean density, and
$\bar\varphi$ is the scalar field value at mean density. 
If the potential $V$ were negligible, one can see that $\phi$ and
$\varphi$ would be proportional to each other, and 
the test particle would accelerate according to an effective
gravitational potential of $\phi 
= \tilde \phi (1 + 2\beta^2)$. More generally, assuming 
the fluctuation of $\varphi$ from $\bar\varphi$ is small, one
can Taylor expand and rewrite the potential terms on the right
as $\sim (a^2/\lambda^2) (\varphi - \bar\varphi)$, where $1/\lambda^2$
is the second derivative of the potential evaluated at $\bar\varphi$. 
One therefore finds $\varphi [1 + a^2/(k^2 \lambda^2)]/(2\beta) = \tilde\phi$
in Fourier space. This means the total effective gravitational potential
$\phi = \tilde\phi \left[1 + 2\beta^2 k^2 /(k^2 +
  a^2/\lambda^2)\right]$. Combining this with the Poisson equation for
$\tilde\phi$ gives us the Yukawa model in Eq. (\ref{yukawa}), if one
makes the identification $\alpha = 2\beta^2$, and $G_N = G(1 +
2\beta^2)$. The Yukawa model we study assumes $\lambda$ is a constant,
but it should evolve in general, depending on the precise form
of the potential $V$.

Note that $\alpha$ is always positive in this sort of 
formulation. The case of a negative $\alpha$ is best regarded
as purely phenomenological.
Note also that depending on the potential
$V$, the scalar-tensor theory could exhibit a chameleon mechanism
on sufficiently small scales, under which $(\varphi - \bar\varphi)/\bar\varphi$
cannot be considered a small perturbation. In this case,
one finds 3 regimes: the largest scales (compared to $\lambda$) 
on which the scalar force is Yukawa suppressed, the intermediate
scales on which the scalar force is operative, and the small scales on
which the chameleon mechanism takes over to suppress the scalar
force. With the chameleon mechanism, there is the additional
possibility of equivalence principle violations, namely that
macroscopic
objects (as opposed to test particles) do not necessarily all fall
at the same rate. See \cite{Hui+09,Hui2010} for further discussions
(note that their $\alpha$ is the same as $\beta$ here).

\section{\label{sec:SC} Spherical collapse}
The Euler equation in an expanding universe is
\begin{eqnarray}
\label{Euler}
\frac{\mathrm{D}\bb v}{\mathrm{D}\eta} = - \frac{a'}{a}\bb v - \del\phi\,,
\end{eqnarray}
where $\eta$ is conformal time, $\bb v$ is the peculiar velocity $\dr \bb x/\dr \eta$
($\bb x$ being comoving position), $a' \equiv \dr a/\dr\eta$, $\del$ is the 
gradient operator with respect to the comoving co-ordinates, and $\phi$ is the
perturbation in the gravitational potential. $\mathrm{D}/\mathrm{D}\eta$
is strictly speaking a material derivative, but, since we are going to follow the
motion of a surface (a spherical shell), we are using Lagrangian
co-ordinates, and 
so can think of this as just a normal time derivative $\dr/\dr\eta$. 
Rewriting this in terms of the physical (proper)
co-ordinate $R$ of a spherical shell, with the scale factor as our time co-ordinate, we find
\begin{equation}
\frac{\dr^2 R}{\dr a^2} = \frac{1}{a^2 H}\left\{ 
  \left( H + a\frac{\dr H}{\dr a} \right)\left( R - a\frac{\dr R}{\dr a} \right) 
 - \frac{1}{H} \nabla_R \phi \right\}\,.
\label{eq:ode}
\end{equation}

We investigate three approaches to calculate the evolution of a
spherically-symmetric perturbation, with an initial top-hat profile.
In the first case, we divide the perturbation into spherical shells and follow the motion of 
each shell under the gravitational force from every other shell. This method includes the
effect of inhomogeneity (i.e. departure from a pure top-hat profile), 
which we might expect to be important because of the violation 
of Gauss's law. The second method approximates the perturbation as a homogeneous sphere at
every time step, and only solves for the motion of the outermost surface. Thirdly,
we use Raychaudhuri's equation to replace Eq. (\ref{eq:ode}),
approximating the problem as local.
In all cases we use a fourth-order Runge-Kutta integrator with 
Cash-Karp co-efficients and adaptive step-size adjustment, similar to that found in 
Press \etal \cite{numrecipes1992}.

The initial conditions for the velocity are 
\[
\frac{\dr R}{\dr a} = \frac{R}{a}\left[ 1 - \frac{\delta}{3(1+\delta)}\right]\,;
\qquad \frac{\dr \delta}{\dr a} = \frac{\delta}{a}\,,
\]
the same as would be used in Newtonian gravity, because these are set early in 
the matter era, when the effects of any modifications to gravity will be negligible.

\subsubsection{Inhomogeneous simulations}
For the inhomogeneous simulations, the force experienced by any shell can be split into
three contributions: that due to the Newtonian $1/r$ part of Eq.~(\ref{eq:potential}),
and those due to the additional Yukawa-type contributions from shells internal and 
external to the point of interest. Integrating the potential measured at $r$ sourced
by a shell of thickness $\dr r'$ centered on $r'$, and taking the gradient with respect
to $r$, we find that the Newtonian-type contribution is
\begin{equation}
\nabla_r\phi_{\mathrm{N}} = \frac{G_N}{1+\alpha} \frac{M_{int} - (4\pi/3)\bar{\rho}(a)r^3}{r^2}\,,
\label{eq:newtgrad}
\end{equation}
where $M_{int}$ is the total mass internal to the proper radius $r$; the Yukawa-type contribution from internal shells is
\begin{align}
\nabla_r\phi_{\mathrm{Y,int}} &= 2 \pi \alpha \frac{G_N}{1+\alpha} \dr r'
\bar{\rho}(a)\delta_i \frac{r'\lambda}{r}\left( \frac{1}{r} + \frac{1}{\lambda}\right)\nonumber\\*
&\quad \times \left\{ \exp{[- (r-r')/\lambda]} - \exp{[-(r + r')/\lambda]} \right\}
\,, 
\label{eq:yukgradint}
\end{align}
where $\delta_i$ is the overdensity of the source shell, and the Yukawa-type contribution from external shells is
\begin{align}
\nabla_r\phi_{\mathrm{Y,ext}} &= - 2\pi \alpha \frac{G_N}{1+\alpha} \dr r'
\bar{\rho}(a)\delta_i \frac{r'\lambda}{r}\nonumber\\* 
&\quad\times\Biggl\{ \left( \frac{1}{\lambda} - \frac{1}{r} \right) \exp{[-(r'-r)/\lambda]}\nonumber\\*
&\qquad + \left( \frac{1}{\lambda} + \frac{1}{r}\right) \exp{[-(r'+r)/\lambda]} \Biggr\}\,.
\label{eq:yukgradext} 
\end{align}

\subsubsection{Homogeneous simulations}
For the homogeneous-sphere simulations, the gradient of the gravitational potential
at the surface of a perturbation of total mass $M$ and radius $R$ is
\begin{align}
\nabla_R\phi &=  \frac{G_N}{1+\alpha} \left( M - \frac{4\pi}{3}\bar{\rho}(a)R^3 \right)
\frac{1}{R^2}\nonumber\\* 
&\quad \times \left\{ 1 + \alpha \left( 1 + \frac{R}{\lambda}\right) \mathcal{F}(R/\lambda) \right\}\,,
\label{eq:homoggrad}
\end{align}
and the form factor \cite{Gibbons:81} is given by 
\[
\mathcal{F}(x) = \frac{3}{x^3} \mathrm{e}^{-x} \left\{ x \cosh(x) - \sinh(x) \right\}\,.
\]

\subsubsection{Local density-field transformation}

Gazta\~{n}aga \& Lobo \cite{Gaztanaga:2000vw} use a combination 
of the continuity and Raychaudhuri
equations to derive a differential equation for the nonlinear evolution of the density
field in Brans-Dicke gravity. 
We adopt their method to our problem.

Taking the divergence of the Euler equation (\ref{Euler}), and
ignoring shear and vorticity, it can be shown that
\begin{eqnarray}
{D\theta \over D\eta} + {1\over 3} \theta^2 + {a' \over a} \theta = -
\nabla^2 \phi \,,
\end{eqnarray}
where it should be kept in mind that $D/D\eta$ is a Lagrangian
derivative, and $\theta \equiv \del\cdot \bb v$. 
Putting this together with the continuity equation $D\delta/D\eta = -
(1+\delta)\theta$, we obtain:
\begin{eqnarray}
\label{localeqt}
{D^2 \delta \over D\eta^2} + {a' \over a} {D\delta \over D\eta}
- {4 \over 3} {1\over 1 + \delta} {D\delta \over D\eta} = (1+\delta)
\nabla^2 \phi
\end{eqnarray}
Note that spatial derivatives in this equation are comoving.
To close this equation, we need to relate $\nabla^2 \phi$ to $\delta$.
Taking the divergence of 
Eq.~(\ref{eq:potential}), and applying the divergence theorem, 
we find $\nabla^2\phi$ located at the center of the spherical top hat
(where shear and vorticity can justifiably be ignored) to be:
\begin{equation}
\nabla^2 \phi = 4\pi \frac{G_N}{1+\alpha} \bar{\rho} a^2 \delta \left\{
1 + \alpha\left(1 + \frac{R}{\lambda}\right) \mathrm{e}^{-R/\lambda} \right\}\,.
\label{eq:raycentre}
\end{equation}
The proper radius $R$ and $\delta$ are related by mass conservation,
i.e. $(1+\delta)R^3/a^3$ is a constant.

\begin{figure}
\includegraphics[width=8.6cm]{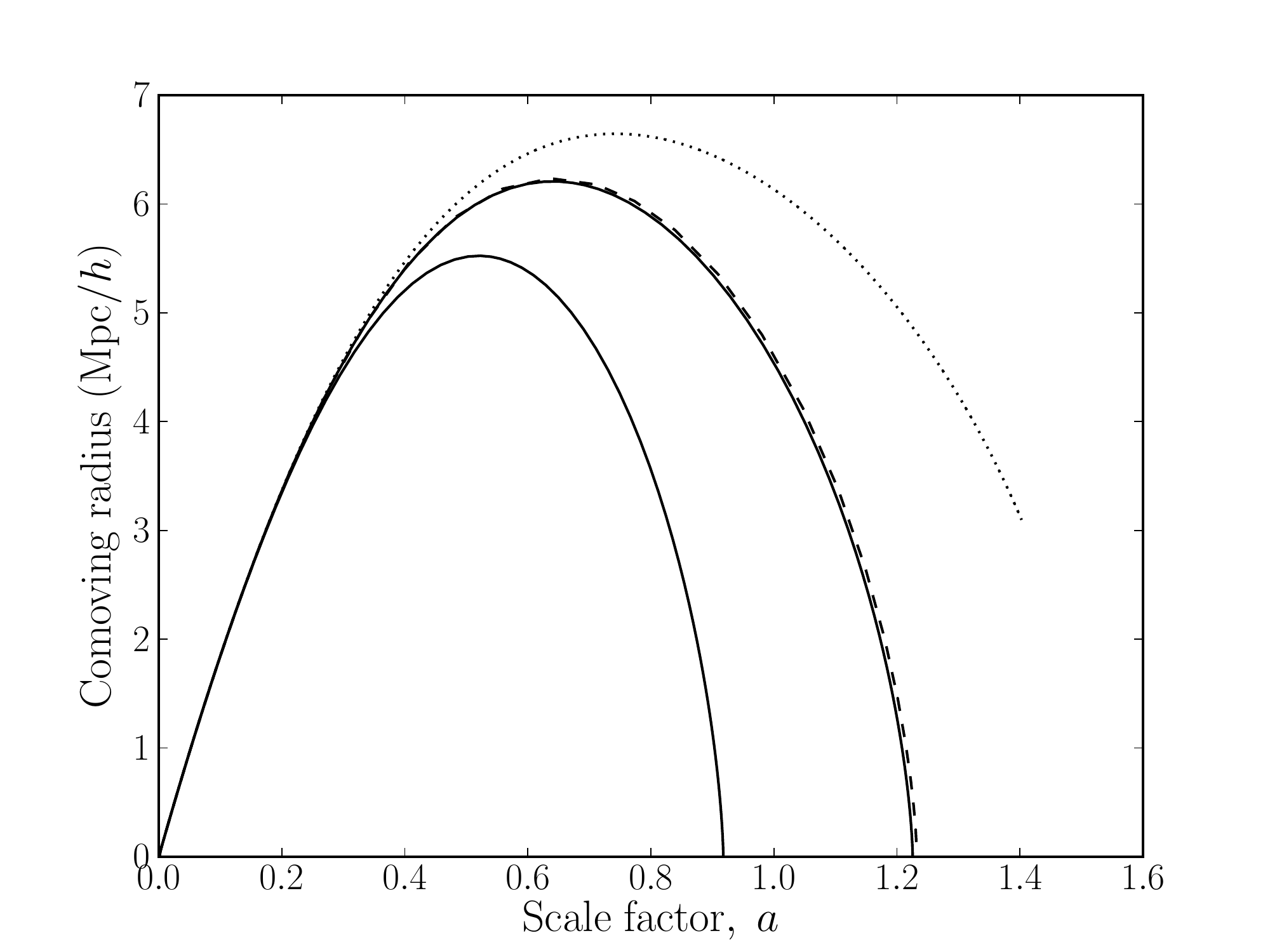}
\caption{\label{fig:compare} Comparison of nonlinear solution methods. 
Lower solid line: Newtonian solution; 
upper solid line: homogeneous simulation; 
dashed line: inhomogeneous simulation; 
dotted line: local transformation (evaluated at center).}
\end{figure}

\subsubsection{Comparison of methods}

The three modified-dynamics methods
give very similar results when the deviation from the Newtonian trajectory is 
small, usually when $R<\lambda$ at all times. However, in more extreme scenarios
the Raychaudhuri equation method overstates the degree to which the gravity modification
changes the trajectory, in comparison to the inhomogeneous sphere method, which we assume is the
most accurate. $R(a)$ trajectories for the three modified gravity solutions are
shown in FIG.~\ref{fig:compare}, along with the associated Newtonian path; 
$\alpha = 5$, $\lambda = 5 \hmpc$.

It appears that the local approximation inherent in the Raychaudhuri method
breaks down when $R \agt \lambda$. However, the homogeneous-sphere 
approximation continues to hold until well beyond the limits
of interest: $M \alt 10^{16}\,\mathrm{M_{\odot}/h}$, $\lambda 
\agt 10\,\hmpc$.

\bibliography{mybib}

\end{document}